\documentclass[usenatbib]{mnras}
\usepackage{graphicx}
\usepackage{dcolumn}
\usepackage{amssymb} 
\usepackage{amsmath}
\usepackage{comment}
\usepackage{ctable}
\usepackage{float}
\usepackage{bm}
\usepackage{float}
\usepackage{epsfig}
\usepackage{epstopdf}
\usepackage{epsf,color}
\usepackage{comment}
\usepackage{aas_macros}
\usepackage{url}
\usepackage{soul}
\usepackage{flushend}
\usepackage{cuted}
\usepackage{color, colortbl}
\newcommand{\cmnt}[1]{}

\usepackage[normalem]{ulem}
\usepackage{color}

\usepackage{tgtermes}

\title[SIDM cross section]{A quantitative comparison between velocity dependent SIDM cross sections constrained by the gravothermal and isothermal models}

\author[S. Yang et al.]{
Shengqi Yang,$^{1,2}$\thanks{E-mail:shengqiy@lanl.gov}
Fangzhou Jiang, $^{1,3}$\thanks{E-mail:fangzhou.jiang@pku.edu.cn}
Andrew Benson,$^{1}$
Yi-Ming Zhong,$^{4,9}$
Charlie Mace,$^{5,6,7}$
\newauthor
Xiaolong Du,$^{8,1}$
Zhichao Carton Zeng,$^{5,6,7}$
Annika H. G. Peter,$^{5,6,7,10}$
Moritz S. Fischer$^{11,12}$\\
$^{1}$Carnegie Observatories, 813 Santa Barbara Street, Pasadena, CA 91101, USA\\
$^{2}$Los Alamos National Laboratory, NM 87545, USA\\
$^{3}$Kavli Institute for Astronomy and Astrophysics, Peking University, Beĳing 100871, People’s Republic of China\\
$^{4}$Kavli Institute for Cosmological Physics, University of Chicago, Chicago, IL 60637, USA\\
$^{5}$Department of Physics, The Ohio State University, 191 W. Woodruff Ave., Columbus, OH 43210, USA\\
$^{6}$Department of Astronomy, The Ohio State University, 140 W. 18th Ave., Columbus, OH 43210, USA\\
$^{7}$Center for Cosmology and AstroParticle Physics, The Ohio State University, 191 W. Woodruff Ave., Columbus, OH 43210, USA\\
$^{8}$Department of Physics and Astronomy, University of California, Los Angeles, CA 90095, USA\\
$^{9}$Department of Physics, City University of Hong Kong, Tat Chee Avenue, Kowloon, Hong Kong SAR, China\\
$^{10}$School of Natural Sciences, Institute for Advanced Study, 1 Einstein Drive, Princeton, NJ 08540\\
$^{11}$Universit\"{a}ts-Sternwarte, Fakult\"{a}tf\"{u}r Physik, Ludwig-Maximilians-Universit\"{a}t M\"{u}nchen, Scheinerstr, 1, D-81679 M\"{u}nchen, Germany\\
$^{12}$Excellence Cluster ORIGINS, Boltzmannstrasse 2, D-85748 Garching, Germany
}

\date{Accepted XXX. Received YYY; in original form ZZZ}

\pubyear{2023}

\begin{document}
\label{firstpage}
\pagerange{\pageref{firstpage}--\pageref{lastpage}}
\maketitle

\begin{abstract}
One necessary step for probing the nature of self-interacting dark matter (SIDM) particles with astrophysical observations is to pin down any possible velocity dependence in the SIDM cross section. Major challenges for achieving this goal include eliminating, or mitigating, the impact of the baryonic components and tidal effects within the dark matter halos of interest---the effects of these processes can be highly degenerate with those of dark matter self-interactions at small scales. In this work we select 9 isolated galaxies and brightest cluster galaxies (BCGs) with baryonic components small enough such that the baryonic gravitational potentials do not significantly influence the halo gravothermal evolution processes. We then constrain the parameters of Rutherford and M{\o}ller scattering cross section models with the measured rotation curves and stellar kinematics through the gravothermal fluid formalism and isothermal method. Cross sections constrained by the two methods are consistent at $1\sigma$ confidence level, but the isothermal method prefers cross sections greater than the gravothermal approach constraints by a factor of $\sim3$.
\end{abstract}




\section{Introduction}
Dark matter is a necessary ingredient to the Standard Model of Cosmology. Yet its particle nature remains mysterious. Self-interacting dark matter (SIDM) is an alternative theory to the standard cold collisionless dark matter (CDM) model. In addition to gravitational interactions, dark matter (DM) particles can also have sizable self-interactions that allow them to scatter off with each other. Such self-interactions are ubiquitous when dark matter is part of so-called ``dark sector'', which consists new  particles that mediated interactions between dark matter-ordinary matter and/or dark matter-dark matter. The SIDM is developed in part to explain discrepancies between predictions of the standard $\Lambda$CDM model and observations at $\sim$kpc scales. More generally speaking, the study of SIDM builds a bridge between astrophysics and particle physics, where the nature of dark matter and dark sectors can be probed from astrophysical observations. Besides the simplest elastic velocity-independent SIDM, astrophysical observations have also been used to explore a wide range of well-motivated SIDM models such as the Yukawa SIDM~\citep{Feng:2009mn,Loeb:2010gj,Tulin:2013teo}, the resonant SIDM~\citep{Chu:2018fzy,Chu:2019awd,Tsai:2020vpi}, the dissipative SIDM~\citep{Boddy:2016bbu,2019PhRvL.123l1102E}, the multi-component inelastic SIDM~\citep{Schutz:2014nka,Blennow:2016gde,Huo:2019yhk}, and the non-abelian SIDM~\citep{Boddy:2014yra, Hochberg:2014dra,Hochberg:2014kqa}.

For a self-gravitating system such as a SIDM halo, the behaviour of SIDM is similar to that of stars in baryonic self-gravitating systems such as the globular clusters \citep{1968MNRAS.138..495L} in which stars undergo strong gravitational scattering events. Since particles with higher energies (larger orbits) have lower velocities (temperature), self-gravitating systems show negative specific heat: hot regions lose heat and become hotter, while cold regions gain heat and get even cooler. As a result, a SIDM halo will never achieve thermal equilibrium. Instead, the halo will go through gravothermal collapse, where the density and temperature of the isothermal cored region at the halo center keep increasing until gravothermal catastrophe. This unique feature of core formation and core collapse provides one possible explanation for the diverse rotation curves observed among galaxies with similar maximum circular velocity (\citealt{2015MNRAS.452.3650O,2020PhRvL.124n1102S,2020JCAP...06..027K,2021MNRAS.503..920C}; see \cite{2018PhR...730....1T} and \cite{2022arXiv220710638A} for a detailed review).\par 
One of the major questions regarding SIDM that must be probed with astrophysical observations is the velocity dependence of the SIDM cross section. This is because different velocity scales correspond to different energy scales of the self-interactions for a fixed dark matter particle mass. Similar to the collider program, probing the behaviours of self-interactions at multiple energy scales could help us understand the underlying  models of particle dark matter that yield self-interaction. If self-interaction is the cause of $\Lambda$CDM-observation discrepancies at small scales, current observations prefer that the SIDM cross section decrease with increasing particle relative velocities \citep{2016PhRvL.116d1302K,2018ApJ...853..109E,2021JCAP...01..024S,2021MNRAS.503..920C,2022MNRAS.510...54A}. Two scattering cases that naturally motivate velocity dependent self-scattering cross sections have been considered in the literature. In the scenario where the two scattering particles are asymmetric and distinguishable $\chi\Bar{\chi}\rightarrow\chi\Bar{\chi}$, only the t-channel contributes to the scattering amplitude. Assuming that the dark matter particle $\chi$ is coupled with a gauge Boson $\phi$ with a coupling strength $g_\chi$, the interaction Lagrangian is $ig_\chi\bar{\chi}\gamma^\mu\chi\phi_\mu$. In the perturbative Born regime $\alpha_\chi m_\chi/m_\phi\ll1$ the differential cross section in the center of momentum frame is (see \cite{2022JCAP...09..077Y} for the derivation):
\begin{equation}
    \dfrac{d\sigma}{d\cos\theta}=\dfrac{\sigma_0\omega^4}{2[\omega^2+v_{12}^2\sin^2(\theta/2)]^2\,.}
\end{equation}
Here $m_\chi$ and $m_\phi$ are the dark matter particle and mediator particle masses, $v_{12}$ is the relative velocity between the two initial scattering particles, $\alpha_\chi=g_\chi^2/(4\pi)$, $\sigma_0=4\pi\alpha_\chi^2/(m_\chi^2\omega^4)$, and $\omega=cm_\phi/m_\chi$. This process is analogous to Rutherford scattering in nuclear physics.\par
In the scenario where the two scattering dark matter particles are indistinguishable $\chi\chi\rightarrow\chi\chi$, both the t-channel and u-channel should be considered for the scattering amplitude calculation. This process is analog to M{\o}ller scattering in the nuclear physics. In the Born regime the center of momentum differential cross section is \citep{2022JCAP...09..077Y}:
\begin{equation}
    \dfrac{d\sigma}{d\cos\theta}=\dfrac{\sigma_0\omega^4[(3\cos^2\theta+1)v_{12}^4+4v_{12}^2\omega^2+4\omega^4]}{(\sin^2\theta v_{12}^4+4v_{12}^2\omega^2+4\omega^4)^2}\,.
\end{equation}


Since SIDM was firstly proposed by \cite{2000PhRvL..84.3760S}, various numerical  (e.g. \cite{2000ApJ...534L.143B,2000ApJ...543..514K,2000ApJ...544L..87Y,2001ApJ...547..574D,2002ApJ...581..777C,2012MNRAS.423.3740V,2013MNRAS.431L..20Z,2013MNRAS.430...81R,2013MNRAS.430..105P,2014MNRAS.444.3684V,2015MNRAS.452.1468F,2015MNRAS.453...29E,2016MNRAS.461..710D,2022MNRAS.513.4845Z}) and semi-analytic  (e.g. \cite{2002ApJ...568..475B,2011MNRAS.415.1125K,2014PhRvL.113b1302K,2015ApJ...804..131P,2016PhRvL.116d1302K,2019PhRvL.123l1102E,2020PhRvD.101f3009N}) tools have been developed to quantitatively model the evolution of halos in SIDM models. Among those, the one-dimensional fluid formalism captures SIDM halo gravothermal evolution with a set of coupled partial differential equations (PDEs). It is suitable for describing isolated and spherically symmetric SIDM halo evolution, and has been shown to provide a good description of results of idealised SIDM N-Body simulations (e.g. \citealt{2011MNRAS.415.1125K,2019PhRvL.123l1102E,2022arXiv220502957Y}). Another advantage is that the fluid formalism can be organized into a scale-free format, so that one set of fluid solutions can be re-scaled for halos with arbitrary masses and scales. While being much more computationally efficient than N-body simulation, the fluid formalism is too expensive for exploring the continuous SIDM cross section parameter space. \cite{2022arXiv220502957Y} (hereafter \defcitealias{2022arXiv220502957Y}{Yang2022}\citetalias{2022arXiv220502957Y}) showed that due to the degeneracy between the SIDM halo thermal conductivity and evolution time, an approximate one-to-one time mapping exists among fluid solutions under different cross section models. Through a non-linear stretching of the halo evolution time axis, a gravothermal fluid solution derived under a constant SIDM cross section can be mapped to solutions for arbitrary velocity dependent cross section models with high accuracy. Such a mapping method further boosts the computational efficiency of the gravothermal fluid formalism and facilitates a thorough exploration of the SIDM cross section parameter space. This approximate universality for the gravothermal fluid solutions is also discussed in \cite{2022arXiv220406568O} and \cite{2022JCAP...09..077Y}.\par 
Another widely adopted semi-analytic SIDM halo evolution model is the isothermal Jeans approach. Unlike the gravothermal fluid formalism that numerically solves the halo heat conduction process, this approach solves the SIDM halo isothermal core with the Jeans-Poisson equation and stitches the cored profile smoothly onto the CDM profile in the halo outskirts. Although it is solving the SIDM halo density profiles in a less ``first-principles'' manner, the isothermal Jeans model is more computationally efficient than the fluid formalism, and the model predictions are in surprisingly good agreement with multiple cosmological SIDM simulations \citep{2021MNRAS.501.4610R,2022arXiv220612425J}. \cite{2022arXiv220612425J} provides the hypothesis that although the isothermal method does not explicitly model the impacts of cosmological mass accretion and mergers on the SIDM halo evolution, the fact that it uses the CDM halo distribution as a boundary condition has implicitly accounted for the halo mass assembling history. This may explain why the isothermal model better reproduces SIDM cosmological simulations than the gravothermal fluid formalism. \par
In this work we combine the fast mapping method proposed by \defcitealias{2022arXiv220502957Y}{Yang2022}\citetalias{2022arXiv220502957Y} with SPARC (Spitzer Photometry \& Accurate Rotation Curves) galaxies \citep{2016AJ....152..157L} and brightest cluster galaxies (BCGs) DM distribution measurements \citep{2013ApJ...765...24N} to jointly fit for angular averaged cross sections under the Rutherford and M{\o}ller scatterings. \cite{2022JCAP...09..077Y} show through N-body simulations that the viscosity cross section: 
\begin{equation}
    \sigma_V=\dfrac{3}{2}\int d\cos\theta\sin^2\theta\dfrac{d\sigma}{d\cos\theta}\,,
\end{equation}
is a good approximation to the results of  angular-dependent cross section models for a halo of $M_{200}\approx10^7 M_\odot$ and $c_{200}\approx20$. More systematic simulation tests are needed to quantify the generality and limitation of this conclusion \citep{2024MNRAS.529.2327F}. Nevertheless, in this work we will consider viscosity cross section:
\begin{equation}\label{eq:Rutherford_vis}
    \sigma_V=\dfrac{6\sigma_0\omega^6}{v_{12}^6}\left[\left(2+\dfrac{v_{12}^2}{\omega^2}\right)\ln\left(1+\dfrac{v_{12}^2}{\omega^2}\right)-\dfrac{2v_{12}^2}{\omega^2}\right]\,,
\end{equation}
for Rutherford scattering, and 
\begin{equation}\label{eq:Moller_vis}
    \begin{split}
        \sigma_V&=\dfrac{3\sigma_0\omega^8}{v_{12}^8+2v_{12}^6\omega^2}\left[2\left(5+\dfrac{5v_{12}^2}{\omega^2}+\dfrac{v_{12}^4}{\omega^4}\right)\ln\left(1+\dfrac{v_{12}^2}{\omega^2}\right)\right.\\
        &\left.-5\left(\dfrac{2v_{12}^2}{\omega^2}+\dfrac{v_{12}^4}{\omega^4}\right)\right]\,,
    \end{split}
\end{equation}
for M{\o}ller scattering. With very different DM halo scales and central velocity dispersion, the SPARC galaxies and BCGs may be able to break the $\sigma_0-\omega$ degeneracy. To avoid uncertainties contributed by the baryonic effects, we select 7 galaxies and 2 BCGs with small baryonic components such that the baryonic gravitational potential at the halo center has negligible impact to the halo gravothermal evolution. We compare the SIDM cross section fitting results constrained through the gravothermal fluid formalism and the isothermal Jeans model, and study how different are these two SIDM models quantitatively.\par 
The plan of this paper is as follows. We review the gravothermal fluid formalism and isothermal Jeans model in Section~\ref{sec:model}. Here we will extend the isothermal method, which is usually assumed to be only valid for describing the SIDM halo core formation process under constant cross section, to the full halo evolution process under arbitrary velocity dependent cross section models. In Section~\ref{sec:targetSelection} we introduce data and data selection criterion used for SIDM cross section model constrain. We introduce the cross section model fitting strategy in Section~\ref{sec:MCMC}, and present the best-fit results in Section~\ref{sec:result}. We finally conclude in Section~\ref{sec:discuss}. Throughout this work we assume cosmological parameters $\Omega_\mathrm{m}=0.307$, $\Omega_\mathrm{b}=0.048$, $\Omega_\Lambda=0.693$, $\sigma_8=0.829$, $n_s=0.96$, and $h=0.678$ \citep{2014A&A...571A..16P,2016MNRAS.457.4340K}.\par

\section{Gravothermal and isothermal models}\label{sec:model}
In this section we briefly review the gravothermal fluid formalism and the isothermal Jeans model. We refer readers to \cite{2002ApJ...568..475B}, \cite{2019PhRvL.123l1102E}, and \defcitealias{2022arXiv220502957Y}{Yang2022}\citetalias{2022arXiv220502957Y} for more detailed review about the gravothermal model, and \cite{2014PhRvL.113b1302K}, and \cite{2022arXiv220612425J} for detailed isothermal Jeans model review.\par
\subsection{Gravothermal fluid formalism}\label{subsec:gravo}
The gravothermal fluid formalism describes an isolated, spherically symmetric, and non-rotating SIDM halo with the following coupled PDEs:
\begin{equation}\label{eq:1}
    \dfrac{\partial M}{\partial r}=4\pi r^2\rho,
\end{equation}
\begin{equation}\label{eq:2}
    \dfrac{\partial(\rho v_\mathrm{rms}^2)}{\partial r}=-\dfrac{G(M+M_\mathrm{b})\rho}{r^2},
\end{equation}
\begin{equation}\label{eq:3}
    \dfrac{L}{4\pi r^2}=-\kappa\dfrac{\partial T}{\partial r},
\end{equation}
\begin{equation}\label{eq:4}
    \dfrac{\rho v_\mathrm{rms}^2}{\gamma-1}\left(\dfrac{\partial}{\partial t}\right)_M\ln\dfrac{v_\mathrm{rms}^2}{\rho^{\gamma-1}}=-\dfrac{1}{4\pi r^2}\dfrac{\partial L}{\partial r}.
\end{equation}
Here $M(r)$ and $M_\mathrm{b}(r)$ denote the DM and baryonic enclosed mass at radius less than $r$, respectively, $\rho(r)$, $v_\mathrm{rms}(r)$, $L(r)$, $\kappa(r)$, and $T(r)=m_\chi v_\mathrm{rms}^2(r)/k_\mathrm{B}$ correspond to SIDM halo density, 1D root-mean-square (rms) velocity averaged over the Maxwell-Boltzmann (MB) distribution, luminosity, thermal conductivity, and temperature at radius $r$, respectively. $k_\mathrm{B}$ is the Boltzmann constant. $v_\mathrm{rms}$ is identical to the 1D velocity dispersion since the halo center is assumed to be at rest. Throughout this work we assume that the DM particles act as monotonic
ideal gas ($\gamma=5/3$).\par 
Two scales characterize the heat conduction caused by DM particle elastic self-scattering: the particle orbital scale height $H=\sqrt{v_\mathrm{rms}^2/(4\pi G\rho)}$ and mean free path $\lambda=m_\chi/(\rho\sigma)$. In the short-mean-free-path (smfp) limit $\lambda\ll H$, DM particles behave like an ideal gas with heat conductivity $\kappa_\mathrm{smfp}=2.1v_\mathrm{rms}k_\mathrm{B}/\sigma$. On the other hand, $\lambda\gg H$ in low density and low cross section systems. In such a long-mean-free-path (lmfp) regime DM particles can orbit for multiple times before experiencing an elastic scattering event. It is conventional to define the heat conductivity $\kappa_\mathrm{lmfp}=0.27\beta nv_\mathrm{rms}^3\sigma k_\mathrm{B}/(\mathrm{G}m_\chi)$, and the total heat conductivity $\kappa=(\kappa_\mathrm{smfp}^{-1}+\kappa_\mathrm{lmfp}^{-1})^{-1}$. Here $\beta$ is
a scaling parameter of order $\mathcal{O}(1)$ that cannot be derived from first-principles. The value of $\beta$ is usually determined through calibrating fluid solutions to N-Body simulations, and is found to be varying in range $0.5-1.5$ \cite[e.g.][]{2019PhRvL.123l1102E,2022arXiv220502957Y,2022arXiv220406568O}.\par
Assuming an NFW SIDM halo initial condition $\rho(r)=\rho_\mathrm{s}/(r/r_\mathrm{s})/(1+r/r_\mathrm{s})^2$, Eq~\ref{eq:1}-\ref{eq:4} can be re-written into scale-free format with $\hat{x}=x/x_0$. Here $x$ is any physical quantity in the PDEs, and $x_0$ is the corresponding unit variable:
\begin{equation}\label{eq:eqset}
    \begin{split}
        \dfrac{\partial \hat{M}}{\partial \hat{r}}&=\hat{\rho}\hat{r}^2\,,\\
        \dfrac{\partial(\hat{\rho}\hat{v}_\mathrm{rms}^2)}{\partial \hat{r}}&=-\dfrac{(\hat{M}+\hat{M}_\mathrm{b})\hat{\rho}}{\hat{r}^2}\,,\\
        \dfrac{\hat{L}}{\hat{r}^2}&=-\hat{\kappa}\dfrac{\partial\hat{v}_\mathrm{rms}^2}{\partial \hat{r}}\,,\\
        \hat{\rho}\hat{v}_\mathrm{rms}^2\left(\dfrac{\partial}{\partial\hat{t}}\right)_{\hat{M}}\ln\dfrac{\hat{v}_\mathrm{rms}^3}{\hat{\rho}}&=-\dfrac{1}{\hat{r}^2}\dfrac{\partial\hat{L}}{\partial\hat{r}}\,.
    \end{split}
\end{equation}
The unit variables are:
\begin{equation}
    \begin{split}
        &r_0=r_\mathrm{s}\,, \rho_0=\rho_\mathrm{s}\,, M_0=4\pi r_\mathrm{s}^3\rho_\mathrm{s}\,,\\ 
        &v_{\mathrm{rms},0}=\sqrt{4\pi G\rho_\mathrm{s}}r_\mathrm{s}\,, t_0=1/\sqrt{4\pi G\rho_\mathrm{s}}\,,\\ &L_0=(4\pi)^{5/2}G^{3/2}\rho_\mathrm{s}^{5/2}r_\mathrm{s}^5\,, (\sigma/m_\chi)_0=1/(\rho_\mathrm{s} r_\mathrm{s})\,.
    \end{split}
\end{equation}
It can be shown that $\hat{\kappa}_\mathrm{lmfp}=0.27\times4\pi\beta\hat{\rho}\hat{v}_\mathrm{rms}^3(\widehat{\sigma/m_\chi})$ and $\hat{\kappa}_\mathrm{smfp}=2.1\hat{v}_\mathrm{rms}/(\widehat{\sigma/m_\chi})$.\par 
Since $\partial\hat{L}/\partial\hat{r}$ determines the halo evolution rate, it is easy to show that $\beta\widehat{(\sigma/m_\chi)}\hat{t}$ are degenerate in the lmfp regime and constant SIDM cross section case. The lmfp gravothermal fluid solution derived under a certain cross section $\widehat{(\sigma/m_\chi)}$ and $\beta$ value can therefore be mapped into solutions under other constant cross sections and $\beta$ parameters through linearly stretching the time axis. \defcitealias{2022arXiv220502957Y}{Yang2022}\citetalias{2022arXiv220502957Y} proves that such a gravothermal solution universality persists for velocity dependent cross section models. This is because at every instance the halo evolution can be captured approximately by a characteristic cross section: 
\begin{equation}\label{eq:thermalsigma}
    \left(\dfrac{\sigma}{m_\chi}\right)_\mathrm{c}=\dfrac{\langle v_1v_2v_{12}^3\sigma(v_{12})/m_\chi\rangle}{\langle v_1v_2v_{12}^3\rangle}\,,
\end{equation}
which is radius independent. Here $v_1$, $v_2$, and $v_{12}$ are velocities of two scattering dark matter particles and their relative velocities. $\langle\rangle$ denotes an integration of the bracketed quantity over a MB velocity distribution, which is characterized by temperature $T$ at the halo center. Since $(\sigma/m_\chi)_\mathrm{c}$ is still time dependent, the one-on-one time mapping relation between gravothermal fluid solution with constant and arbitrary velocity dependent cross section models is non-linear.\par
\subsection{Isothermal model}\label{subsec:iso}
Unlike the gravothermal fluid formalism that numerically solves the entire SIDM halo evolution process over a sequence of time steps, the isothermal Jeans model starts by analytically solving for the isothermal cored profile through the Jeans equation Eq~\ref{eq:2} and Poisson equation:
\begin{equation}
    \dfrac{1}{r^2}\dfrac{d}{dr}\left(r^2\dfrac{d\Phi}{dr}\right)=4\pi G(\rho+\rho_\mathrm{b})\,.
\end{equation}
Here $\rho_\mathrm{b}$ is some parameterized baryon density profile such as the Hernquist, and is usually assumed to be static. The density and velocity dispersion of the cored region $\{\rho_\mathrm{c},v_\mathrm{rms,c}\}$ are then adjusted such that the cored density profile joins smoothly onto the CDM halo outskirts at a characteristic radius $r_1$, outside of which SIDM particles have, on average, experienced less than one collision:
\begin{equation}\label{eq:r1}
    \dfrac{4}{\sqrt{\pi}}\rho(r_1)v_\mathrm{rms}(r_1)\dfrac{\sigma}{m_\chi}=\dfrac{1}{t}\,.
\end{equation}
In Eq~\ref{eq:r1} $\sigma$ and $t$ are degenerate. Now consider a velocity dependent cross section model---$r_1$ should be estimated through:
\begin{equation}\label{eq:r1_v}
    \dfrac{4}{\pi}\rho(r_1)v_\mathrm{rms}(r_1)\int_0^t\left(\dfrac{\sigma}{m_\chi}\right)_\mathrm{c}(t^\prime)\mathrm{d}t^\prime=1\,.
\end{equation}
We can move $\rho(r_1)v_\mathrm{rms}(r_1)$ outside the time integration here because at $r\geq r_1$ the halo density and velocity dispersion radial profiles remain unchanged from the NFW initial condition, i.e. they do not evolve with time due to a lack of DM self-interactions. The thermal averaged cross section $(\sigma/m_\chi)_\mathrm{c}$ is defined in Eq~\ref{eq:thermalsigma}. Here we have implicitly assumed that the gravothermal and isothermal method share a similar universality in halo evolution. More specifically, Eq~\ref{eq:thermalsigma} is derived from the thermal average of heat flux defined in the gravothermal formalism. It is impossible to derive $(\sigma/m_\chi)_\mathrm{c}$ for the isothermal model in a self-consistent way since the isothermal model does not involving solving for the heat flux through the halo. Here, based on the similarity between these two methods, we assume $(\sigma/m_\chi)_\mathrm{c}$ suggested by the gravothermal formalism also applies to the isothermal model.  Eq~\ref{eq:r1_v} shows that for a SIDM halo with fixed CDM boundary condition and static baryon distribution, the isothermal solution is determined solely by the integral $\int_0^r(\sigma/m_\chi)_\mathrm{c}(t^\prime)\mathrm{d}t^\prime$. Once a set of isothermal solutions is derived over a sequence of finely gridded halo evolution time $t$ under a constant SIDM cross section $\sigma/m_\chi$, given a velocity dependent SIDM cross section model one can compute $(\sigma/m_\chi)_\mathrm{c}$ during each time step $\mathrm{d}t$, and rescale the time interval into $\mathrm{d}t^\prime$ such that $(\sigma/m_\chi) \mathrm{d}t=(\sigma/m_\chi)_\mathrm{c}(t^\prime)\mathrm{d}t^\prime$, thus deriving the isothermal solutions under arbitrary velocity dependent cross section models. In other words, the time mapping method introduced by~ \defcitealias{2022arXiv220502957Y}{Yang2022}\citetalias{2022arXiv220502957Y} is also applicable to the isothermal solutions.\par
\begin{figure}
    \centering
    \includegraphics[width=0.45\textwidth]{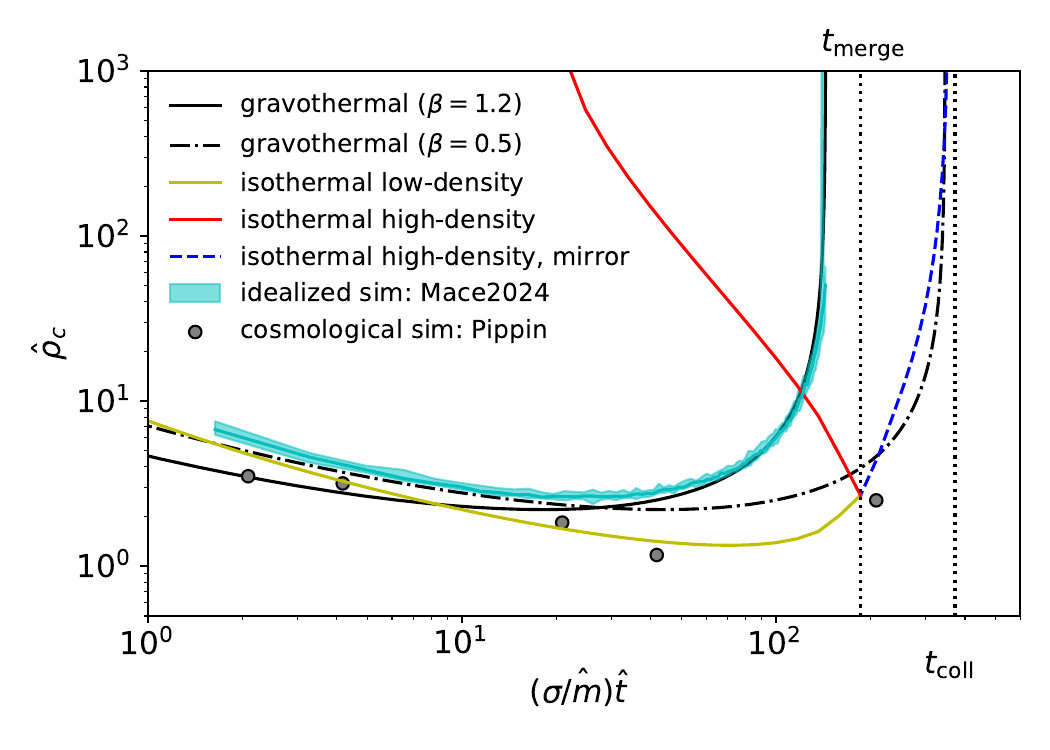}
    \caption{Scale-free SIDM halo central density versus halo evolution time comparison between the gravothermal fluid formalism, the isothermal model, and numerical simulations \protect\citep{2015MNRAS.453...29E,2024arXiv240201604M}. The yellow and red curves show the low and high density isothermal solutions, merging at $t_\mathrm{merge}$. Flipping the high density solution according to $t_\mathrm{merge}$ gives the isothermal solutions during the core collapse stage, which is shown by the blue dashed curve.}\label{fig:gravo-iso}
\end{figure}

We compare the scale-free DM-only halo central density evolution given by the gravothermal and isothermal models in Figure~\ref{fig:gravo-iso}, together with idealized SIDM N-Body and cosmological simulation results. Throughout this work we will assume that halos show NFW density profiles under the CDM scenario, that is, we will always assume a NFW profile as the initial (boundary) condition for the gravothermal (isothermal) model. The cyan band in Figure~\ref{fig:gravo-iso} shows the maximum, median, and minimum halo central density versus evolution time for 10 idealized N-Body realizations of a SIDM halo of mass $M_{200}=10^{10.5}M_\odot$, concentration $c_{200}=10$, and particle number $10^6$ \citep{2024arXiv240201604M}. The gravothermal solution is in excellent agreement with this set of idealized N-Body simulation results with $\beta=1.2$. On the other hand, the isothermal solutions are in better agreement with cosmological SIDM simulation results \citep{2015MNRAS.453...29E}.\par
\begin{figure*}
    \centering
    \includegraphics[width=0.45\textwidth]{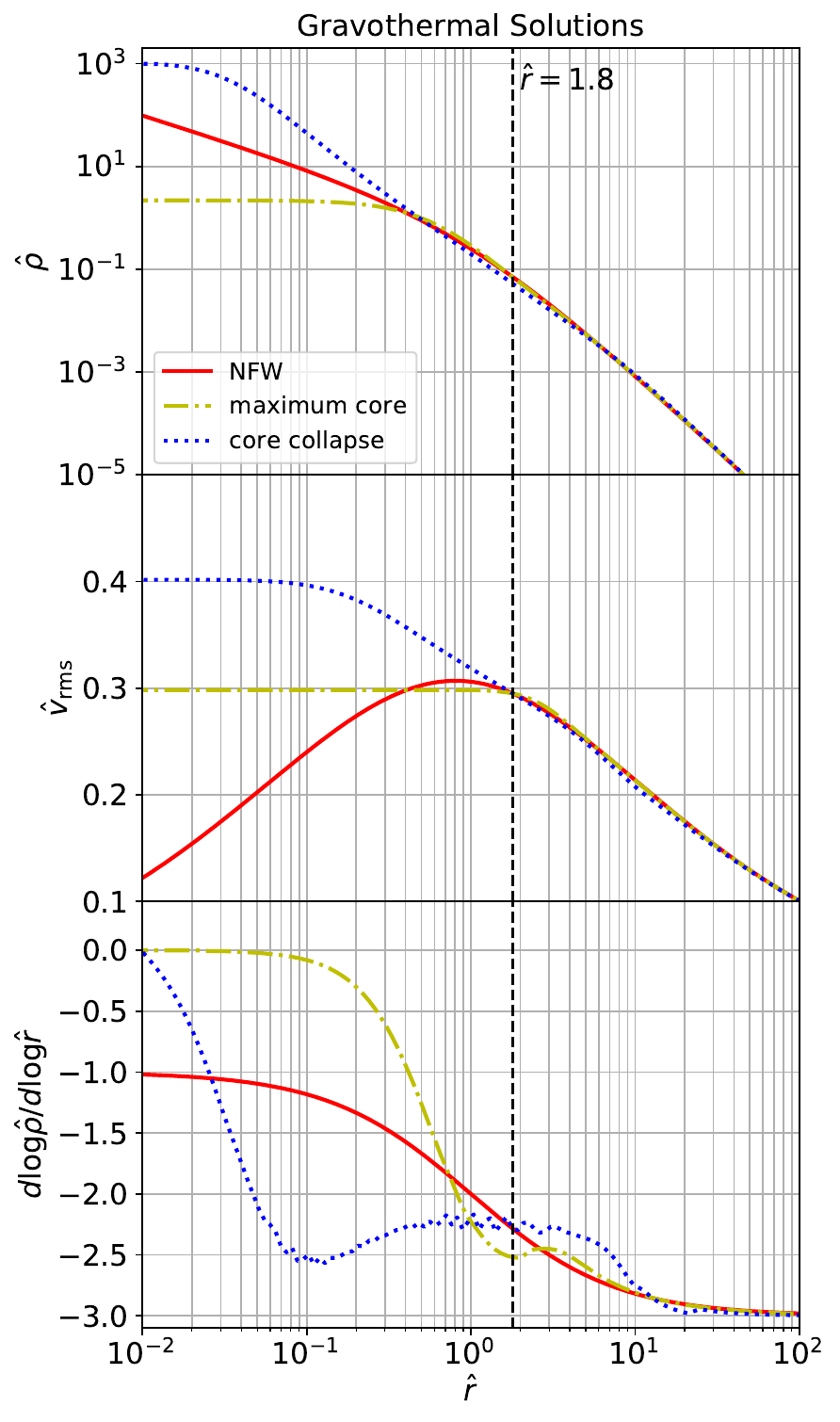}
    \includegraphics[width=0.45\textwidth]{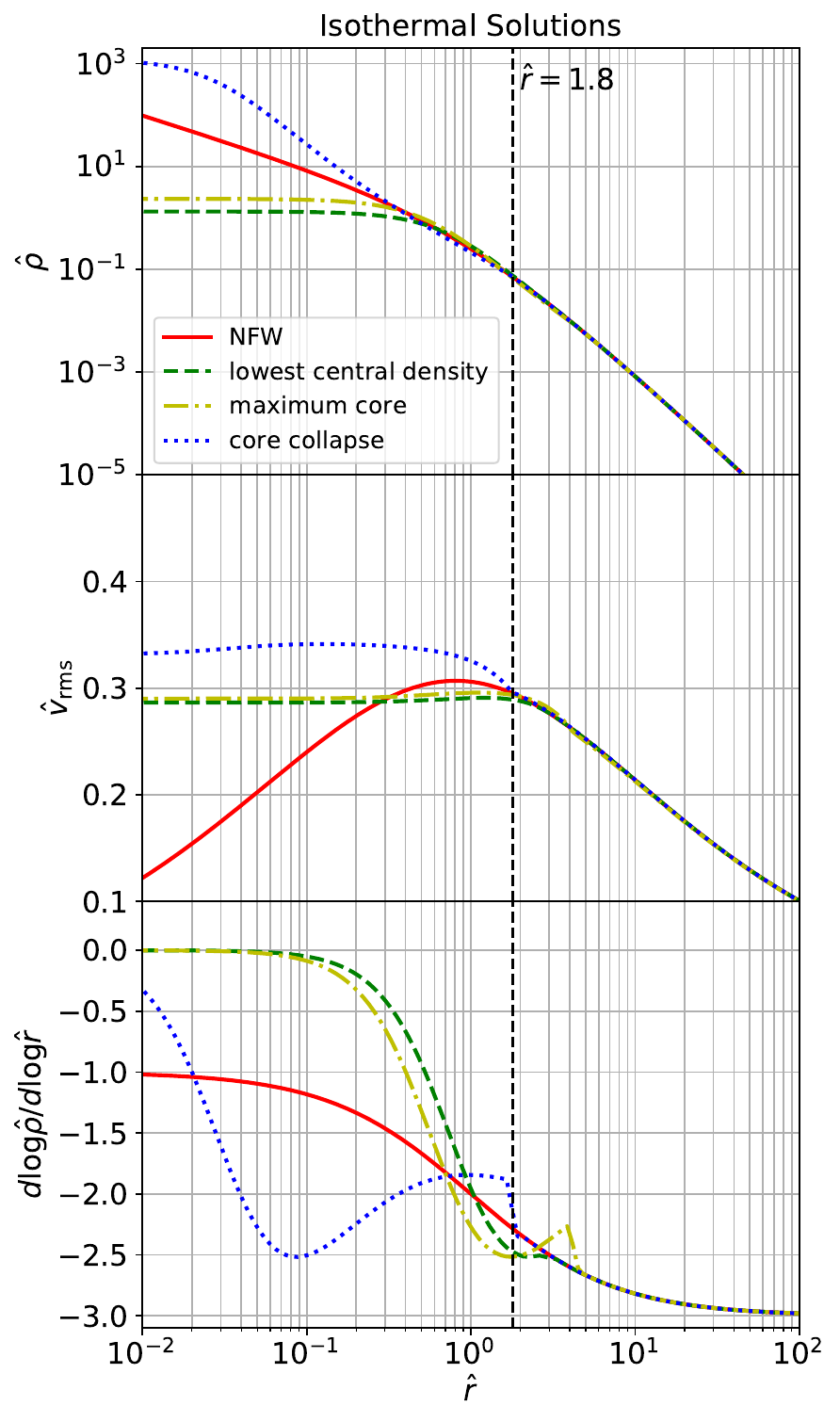}
    \caption{Scale-free SIDM halo density profiles (top row), velocity dispersion radial profiles (middle row), and the density profile slope (bottom row) at the initial condition (red), maximum cored instance (yellow dashdot), and deep core collapse stage (blue dotted) given by the gravothermal fluid formalism (left column) and isothermal model (right column). The halo density profiles and velocity dispersion profiles are always close to the NFW initial condition at $\hat{r}\gtrsim1.8$, marked by the black dashed vertical line. The halo maximum cored instance is slightly delayed from the lowest central density instance (green dashed line) for the isothermal solutions, while those two instances are identical for the gravothermal solutions.}\label{fig:gravothermalSolutions}
\end{figure*}

 
Figure~\ref{fig:gravo-iso} shows that during the SIDM halo core formation process where the halo central density decreases with time, there are two sets of $\{\rho_\mathrm{c},v_\mathrm{rms,c}\}$ that stitch the isothermal Jeans solution smoothly to the NFW outskirts at $r_1$. The lower density solution (yellow curve) is treated as the valid modeling result, while the higher density solution (red curve) is usually discarded since its physical meaning is unclear. As the halo evolution time increases, these two isothermal solutions get closer to each other, and finally merge at $t_\mathrm{merge}$. At $t>t_\mathrm{merge}$ there is no $\{\rho_\mathrm{c},v_\mathrm{rms,c}\}$ that stitches the isothermal cored profile with the NFW outskirt, and the isothermal Jeans model is generally thought to be no longer applicable. We hypothesize the high density solutions provide the halo density profiles in the core-collapsing process. Specifically, we generalize Eq~\ref{eq:r1} into:
\begin{equation}\label{eq:r1_gen}
    \dfrac{4}{\sqrt{\pi}}\rho(r_1)v_\mathrm{rms}(r_1)\dfrac{\sigma}{m_\chi}=\mathrm{max}\left[\dfrac{1}{t},\dfrac{1}{t_\mathrm{coll}-t}\right]\,.
\end{equation}
Here $t_\mathrm{coll}$ is the time when the SIDM halo collapses. When $t<t_\mathrm{coll}/2=t_\mathrm{merge}$, Eq~\ref{eq:r1_gen} reduces to Eq~\ref{eq:r1}. The SIDM halo distributions at $r>r_1$ should be identical to NFW because DM particles self-scatter is statistically unimportant at the halo outskirt. On the other hand, at $t>t_\mathrm{coll}/2$ the halo density profile at a certain radius will show negligible time evolution if the DM particle self-interaction probability is less than 1 during the rest of the halo lifetime $t_\mathrm{coll}-t$. We notice that at $t>t_\mathrm{coll}/2$ it is no longer exact to describe the instantaneous halo outskirt distribution as NFW because DM particle self-scattering was statistically important in the past and had altered the halo density and velocity dispersion profile shapes. However, NFW is still a good approximation at large radii. In Figure~\ref{fig:gravothermalSolutions} we compare the scale-free halo density profiles and velocity dispersion radial profiles in the NFW initial conditions, at the instant of maximum core radius, and deep into the core collapsing phase given by the gravothermal and isothermal models. Here we have assumed that the halo is DM-only ($M_\mathrm{b}=\rho_\mathrm{b}=0$) and the cross section is constant. For both the gravothermal and isothermal solutions, the halo velocity dispersion profiles are very close, although not identical, to the NFW profile at $\hat{r}\gtrsim1.8$. We have checked that for the DM-only case, we can only find numerically stable high density solutions if $\hat{r}_1\gtrsim1.8$. The gravothermal solution velocity dispersion at $\hat{r}\geq1.8$ is only different from the NFW profile by less than 3\% at core collapse, therefore it is a good approximation to stitch the isothermal cored profile to the NFW profile at $t>t_\mathrm{merge}$ for modeling the SIDM halo evolution. Since it is the rest of the halo lifetime $t_\mathrm{coll}-t$ that matters during the halo core collapsing phase, one will need to mirror the high density solution according to $t_\mathrm{merge}$ to derive the halo density profiles at $t>t_\mathrm{merge}$ (blue dashed curve in Figure~\ref{fig:gravo-iso}). The isothermal model can be played back in time because there is no such a notion of ``the arrow of time". The evolution could back played from the $t_\mathrm{coll}$ to an earlier time and the same formalism to compute the density profile applies. Note that this is different from the fluid formalism, where the entropy ($\ln(v_\mathrm{rms}^2/\rho^{\gamma-1})$) is moving toward an ever larger value as time evolves. Therefore, the fluid formalism has the ``arrow of time". In Figure~\ref{fig:gravo-iso} we present the gravothermal solution for $\beta=0.5$, which collapses at the same time as the extended isothermal solution, to better compare the time evolution between these two methods.\par
Figure~\ref{fig:gravo-iso} also shows that the halo central densities given by the DM-only gravothermal and isothermal models differ by a factor of $\sim2$ at the instant when the halo central density reaches minimum. This difference is caused by the fact that the gravothermal fluid formalism treats the NFW profile as the initial condition at $t=0$, while the isothermal model treats the NFW profile as a boundary condition at $r=r_1$. From a more quantitative view, the SIDM halo core size $r_1$ first increases and then decreases with time in both gravothermal and isothermal models. In the gravothermal fluid formalism the halo core extends during the core formation process, and contracts immediately after the instance when the halo central density reaches its minimum. In the isothermal model, however, the halo core size keeps increasing for a while after the halo central density reaches its minimum. In other words, there is a slight mismatch between the halo central density contraction and core size contraction in the isothermal model.  
We present the scale-free isothermal solutions at the minimum halo central density instance and the maximum cored instance in Figure~\ref{fig:gravothermalSolutions} right panel. This slight mismatch leads to the major difference between the gravothermal and isothermal solutions. It is unclear whether this mismatch is physical or a coincidence. Another difference is that a greater radial range of the gravothermal density profile is altered by the DM self-interaction and becomes different from NFW as the halo evolution time increases. However, the isothermal profiles are only different from the NFW at $r<r_1$ by construction. We compare the gravothermal and isothermal density profile slopes at the bottom row of Figure~\ref{fig:gravothermalSolutions} to highlight their slight differences. Besides the above two points, the gravothermal and isothermal solutions are very similar to each other throughout the halo evolution.\par
The middle right panel of Figure~\ref{fig:gravothermalSolutions} shows that the SIDM halo velocity dispersion radial profile at $r\leq r_1$ is not strictly a constant. This is because the isothermal model iteratively solves for a trial solution of the Jeans-Poisson equation for the inner halo until the inner-halo's SIDM profile ``stitches" smoothly to the outer boundary condition at $r_1$ set by the CDM profile. Here, ``stitch" refers to minimizing the difference between the outer CDM profile and the inner SIDM profile in density and enclosed mass \citep{2014PhRvL.113b1302K, 2016PhRvL.116d1302K, 2021MNRAS.501.4610R}. This minimization process has some tolerance, where the objective function is a combination of the relative differences in mass and density, but has no criteria on the velocity dispersion profile. Hence, there is no guarantee in the stitching process that the original assumption of a constant velocity dispersion must strictly hold. With these caveats acknowledged, we note that the profile within $r_1$ is still highly isothermal---the only exception is for the last 10 percent of the radius range near $r_1$, where the smooth asymptotic behavior imposed by the ``stitching" is expected.\par

\par

\par

\section{Data}\label{sec:targetSelection}
DM distribution measurements for systems of at least two very different scales are required to constrain the two parameter cross section models Eq~\ref{eq:Rutherford_vis} and Eq~\ref{eq:Moller_vis}. In this work we select isolated and baryon-poor systems among 175 SPARC galaxies \citep{2016AJ....152..157L} and 7 BCGs \citep{2013ApJ...765...24N} for joint SIDM cross section analysis. SPARC provides a representative sample of disk galaxies in the nearby Universe, with DM halo mass distribution peaks at $M_{200}=10^{10}-10^{12}M_\odot$. Assuming each SPARC galaxy host halo is described by an NFW density profile, the distribution of the maximum velocity dispersion peaks at $30-100$ km/s. The BCG halo mass distribution covers a much higher range $M_{200}=10^{14.5}-10^{15.4}M_\odot$, corresponding to NFW maximum velocity dispersion of $600-1000$ km/s. We do not include bright Milky Way dwarf spheroidal galaxies in the data set in order to avoid modeling tidal effects.\par
\begin{figure*}
    \centering
    \includegraphics[width=0.45\textwidth]{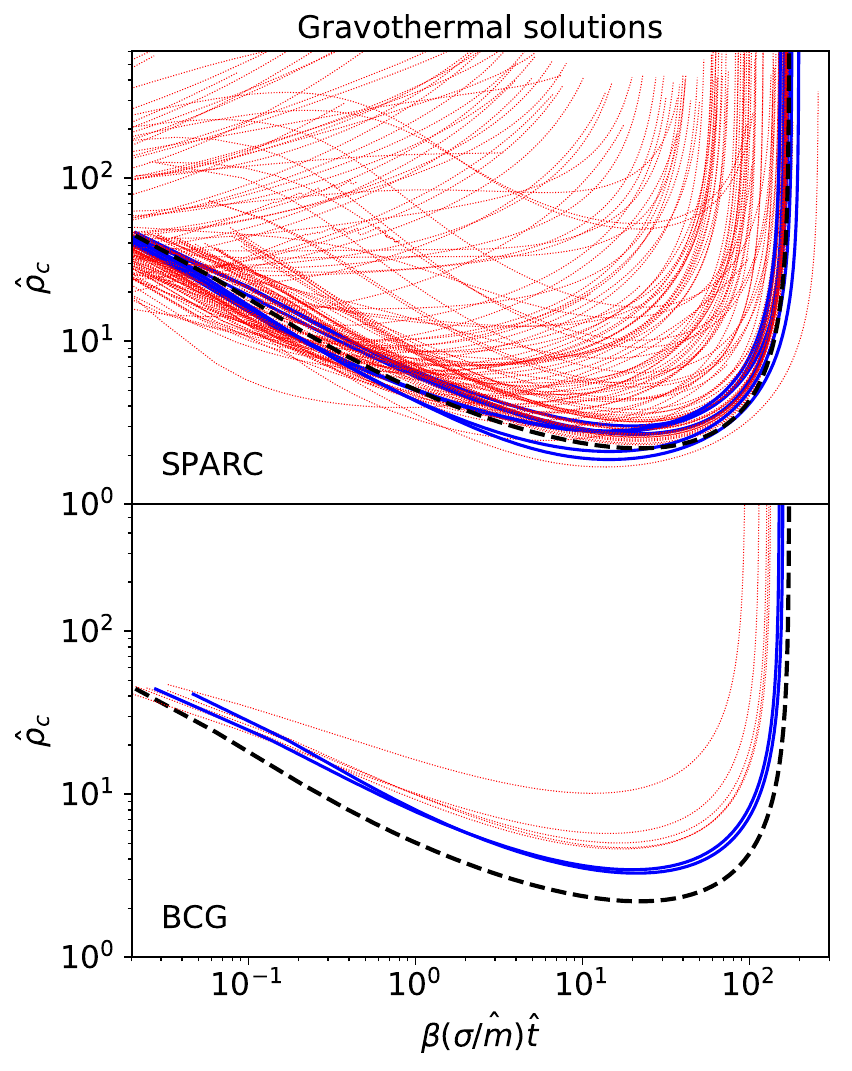}
    \includegraphics[width=0.45\textwidth]{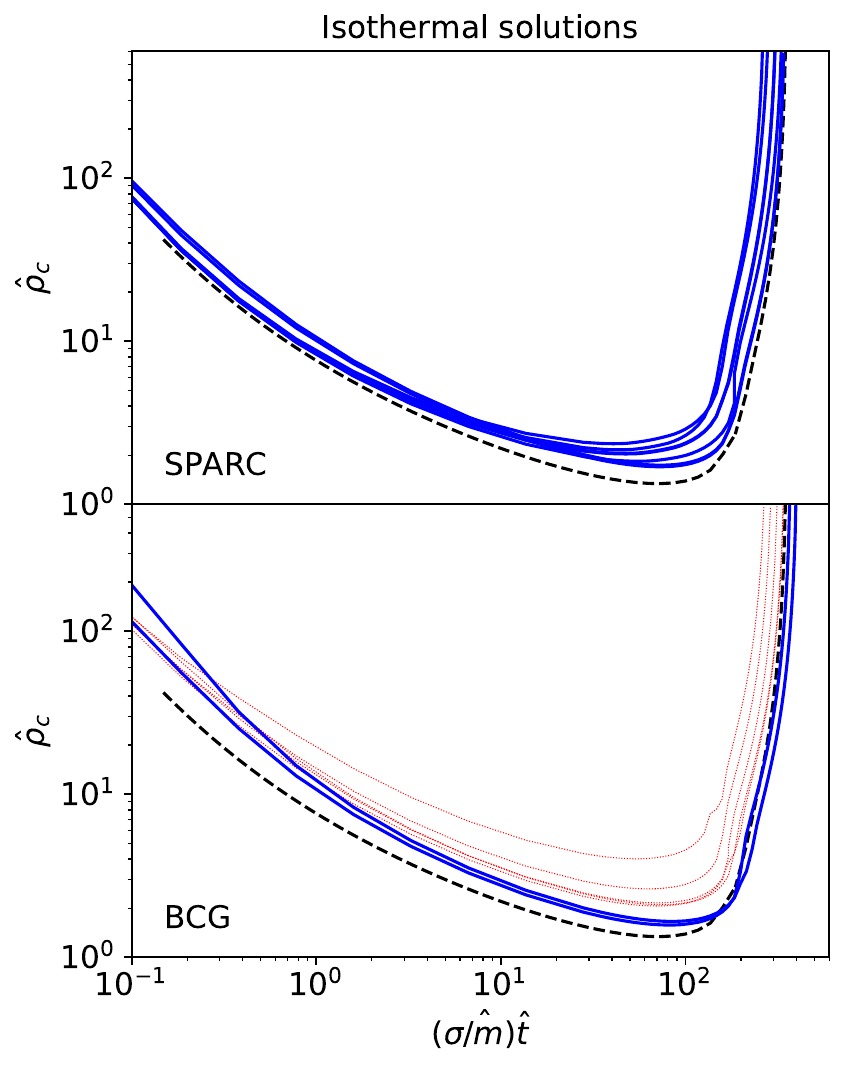}
    \caption{Scale-free halo central density versus evolution time relation with (red dotted) and without (black dashed) the presence of a baryonic gravitational potential at the halo center. Halo central density evolution is highlighted in blue for the selected baryon-poor systems. \textit{Upper left}: Fluid solutions for 142 SPARC galaxies. \textit{Bottom left}: Fluid solutions for 7 BCGs. The second column shows isothermal solutions for the selected SPARC galaxies and the 7 BCGs.}\label{fig:rhoct}
\end{figure*}\par
The recent study of \cite{2023MNRAS.526..758Z} adds various static gravitational potentials contributed by the baryonic component to the halo center and shows that the gravothermal solution is in good agreement with N-Body simulations for isolated SIDM halos with central baryons after $\beta$ calibration. \cite{2022arXiv220612425J} shows that although the isothermal Jeans model does not account for cosmological mass accretion and merger, it is in excellent agreement with cosmological DM-only SIDM simulation \citep{2015MNRAS.453...29E} and cosmological zoom-in simulations \citep{2021MNRAS.507..720S}. These works provide evidence that the halo adiabatic contraction and time variation of the baryonic component gravitational potential may cause only limited impact on the SIDM halo gravothermal evolution, and therefore do not significantly undermine the reliability of the gravothermal and isothermal models. Nevertheless, in this work we still limit our data set to systems with small baryonic components. As such, our results are conservative with regard to possible baryonic effects. Also, the strength of time mapping method of \defcitealias{2022arXiv220502957Y}{Yang2022}\citetalias{2022arXiv220502957Y} adopted in this work is that the gravothermal and isothermal models can be written into a scale-free format, so that the modeling results can be quickly rescaled for halos with arbitrary NFW parameters $\{\rho_\mathrm{s},r_\mathrm{s}\}$ and SIDM cross sections. This is true for the DM-only case. However, the presence of baryons brings additional scales into the problem. Consider, for example, the gravothermal fluid formalism. The SPARC galaxy rotation curves are mostly measured at $r<r_\mathrm{s}$, where the NFW initial conditions are degenerate in the combination $\rho_\mathrm{s}r_\mathrm{s}$. By fitting gravothermal solution derived from a fixed $\hat{M}_\mathrm{b}(\hat{r})$ profile, the physical baryon enclosed mass profile $M_\mathrm{b}=4\pi\rho_\mathrm{s}r_\mathrm{s}^3\hat{M}_\mathrm{b}$ varies when the $\{\rho_\mathrm{s},r_\mathrm{s}\}$ parameter space is explored. Through limiting the dataset to baryon-poor systems, we can partially avoid this inconsistency because the baryonic gravitational potential itself has negligible impact on the gravothermal solutions. This is less of a problem among BCGs, for which the DM distributions are measured over a much greater radial extent $r>r_\mathrm{s}$. At large radii the halo NFW initial conditions are degenerate in the combination $\rho_\mathrm{s}r_\mathrm{s}^3$, leaving $M_\mathrm{b}$ invariant under the $\{\rho_\mathrm{s},r_\mathrm{s}\}$ variation.\par
To select baryon-poor galaxies from the 175 SPARC samples, we first fit the mapped gravothermal fluid solutions to the SPARC DM rotation curves following method introduced in \defcitealias{2022arXiv220502957Y}{Yang2022}\citetalias{2022arXiv220502957Y} to constrain the NFW parameters of each galaxy. In this step we assume a halo evolution time $t=10$ Gyr and $\beta=0.5$. Due to the $\beta(\sigma/m_\chi)t$ degeneracy in Eq~\ref{eq:4} in the lmfp regime, the specific choices of $t$ and $\beta$ values only influence the constraints obtained on the cross section, $\sigma_0/m_\chi$, and have no impact on the $\{\rho_\mathrm{s},r_\mathrm{s}\}$ constraints for each galaxy. Since our model contains 4 free parameters $\{\sigma_0/m_\chi,\omega,\rho_\mathrm{s},r_\mathrm{s}\}$, we consider only systems with non-negative DM rotation curves measured in at least 5 radial bins. This selection criterion reduces the sample size from 175 to 142. With the best fit $\{\rho_\mathrm{s},r_\mathrm{s}\}$ of each galaxy, we then add baryonic enclosed mass to the scale-free Jeans equation. We estimate the enclosed baryonic mass as a function of radius $M_\mathrm{b}(r)$ for each SPARC galaxy through linearly interpolating the $M_\mathrm{b}$ given by the gas, disk, and bulge rotation curves: $M_\mathrm{b}(r_\mathrm{RC})=[v_\mathrm{gas}^2(r_\mathrm{RC})+v_\mathrm{disk}^2(r_\mathrm{RC})+v_\mathrm{bulge}^2(r_\mathrm{RC})]r_\mathrm{RC}/G$. Here $r_\mathrm{RC}$ denotes the rotation curve radial bins. We then re-solve the gravothermal fluid formalism for all the 142 galaxies at fixed $\beta$ and constant SIDM cross section, and compare the halo central density time evolution to the DM-only fluid solutions. The comparison results are presented in Figure~\ref{fig:rhoct} upper left panel. We show the scale-free halo central density time evolution $\hat{\rho}_\mathrm{c}(\beta\widehat{(\sigma/m_\chi)}\hat{t})$ given by the gravothermal fluid formalism without any baryonic component ($\hat{M}_\mathrm{b}=0$) as the black dashed curve, while the fluid solutions for the 142 SPARC galaxies with different discs are shown as red dotted lines. We find 20 SPARC galaxies containing sufficiently small discs such that their fluid solutions are not significantly altered by the baryonic gravitational potential \footnote{We define systems with a small baryonic component as those with gravothermal solutions with collapse time, maximum cored instance, minimum halo central density, and the first time step $\mathrm{d}\hat{t}_1$ less than 15\%, 10\%, 60\%, and 200\% different than those of the DM-only gravothermal solution.}. Finally, we exclude all SPARC galaxies with best fit $r_\mathrm{s}>100$ kpc from the data set. This is because we have assumed zero baryon density beyond the range for which the rotation curve is measured. The scale-free baryon enclosed mass $\hat{M}_\mathrm{b}$ can therefore be small if $r_\mathrm{s}$ is much greater than the outer extent of the rotation curve. There are 7 galaxies that satisfy all the above selection criterion: D564-8, NGC3741, F568-V1, UGC00731, UGC07608, F563-1, and UGC05764. The selected baryon-poor galaxies are highlighted in blue in the figure. We notice that this selection could, in principle, leave us with a SPARC galaxy sample that is somehow biased in the formation processes. For example, it is possible that the selected systems are baryon-poor because they formed much later than most of the galaxies of the similar mass. If this is the case, we will be systematically overestimating the halo evolution time and underestimating the SIDM cross sections for the selected SPARC galaxies.\par
We adopt best fit $\{\rho_\mathrm{s},r_\mathrm{s}\}$ values and dPIE baryonic density profiles reported in \cite{2013ApJ...765...24N} to derive fluid solutions with baryonic gravitational potentials for 7 BCGs. We find that 2 BCGs with the greatest $r_\mathrm{s}$ (A2667 and A2390) show fluid solutions similar to the no-baryon case. The $\hat{\rho}_\mathrm{c}(\beta\widehat{(\sigma/m_\chi)}\hat{t})$ comparisons are presented in Figure~\ref{fig:rhoct} bottom left panel.\par
In this work we use the publicly available isothermal model \cite{2022arXiv220612425J} for isothermal solution derivations. We fit the baryon rotation curve for all SPARC galaxies and BCGs with the Hernquist profile $\rho_\mathrm{b}(r)=M_\mathrm{b}r_\mathrm{s,b}/(2\pi r)/(r+r_\mathrm{s,b})^3$. Here $M_\mathrm{b}$ is the total baryon mass. To ensure physically meaningful fitting results for SPARC galaxies, we set $M_\mathrm{b}$ upper bound as the maximum value between $0.1M_{200}$ and the maximum baryonic mass suggested by the rotation curve at the maximum radial bin. We compare the best fit Hernquist profile with measurements for the selected SPARC galaxies and all BCGs in Appendix~\ref{apdx:Hernquist}. Together with the best fit $\{\rho_\mathrm{s},r_\mathrm{s}\}$ parameters derived previously, we are able to solve for the isothermal profiles for the selected systems over a grid of halo evolution times. As discussed in Section~\ref{sec:model}, we mirror the high density isothermal solutions according to $t_\mathrm{merge}$ and obtain the halo information during core collapse. Figure~\ref{fig:rhoct} second column shows comparisons among isothermal solutions with and without a static baryon gravitational potential present at the halo center. The top row presents isothermal solutions for the 7 selected SPARC galaxies, while the bottom row corresponds to the 7 BCGs. We find the presence of a baryonic component does not significantly alter the isothermal solutions for the selected galaxies and BCGs.\par

\begin{figure}
    \centering
    \includegraphics[width=0.45\textwidth]{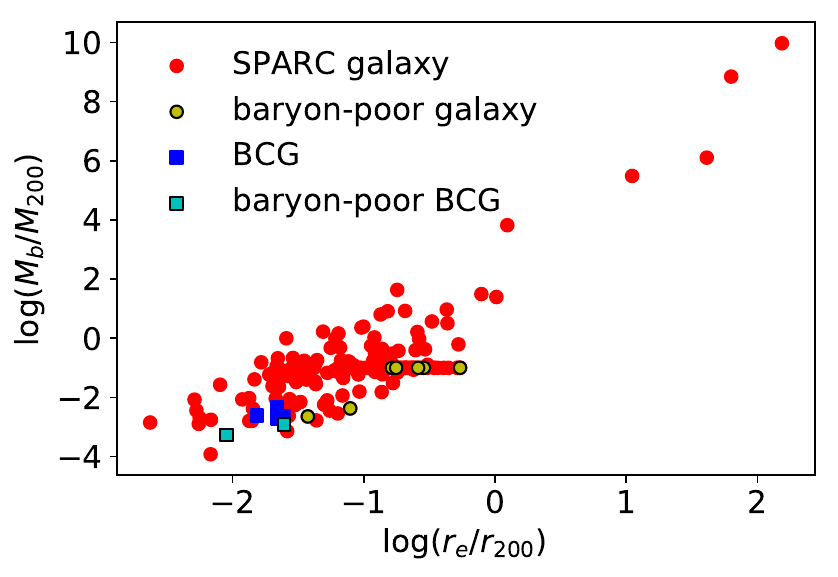}
    \caption{Best-fit baryon versus DM mass ratios and scale ratios for all SPARC galaxies and BCGs, as well as the selected baryon-poor systems. Less massive or more extended baryonic components have less influence on the SIDM halo evolution, therefore the selected systems are found at the bottom of the $M_\mathrm{b}/M_{200}$ versus $r_e/r_{200}$ band.}\label{fig:MR}
\end{figure}

It is challenging to provide a quantitative selection criterion of the baryon-poor systems without solving the gravothermal fluid formalism numerically. However, we do notice that a less massive or more extended baryonic distribution results in less impact on the gravothermal solution. To illustrate this, we present $M_\mathrm{b}/M_{200}$ versus $r_\mathrm{e}/r_{200}$ for the 142 SPARC galaxies (red points) and 7 BCGs (blue points) in Figure~\ref{fig:MR}. Here $M_\mathrm{b}$ and $r_\mathrm{s,b}$ are the best fit Hernquist parameters. $r_\mathrm{e}=1.815r_\mathrm{s,b}$ is the effective radius, and $M_{200}$ and $r_{200}$ are the best fit DM halo virial mass and radius, defined with density contrast $\Delta_\mathrm{vir} = 200$ with respect to the critical density. Figure~\ref{fig:MR} shows that at fixed $r_\mathrm{e}/r_{200}$, the selected baryon-poor systems tend to show the lowest $M_\mathrm{b}/M_{200}$. On the other hand, at fixed $M_\mathrm{b}/M_{200}$ the baryon-poor systems tend to host extended baryon distributions, showing some of the highest $r_\mathrm{e}/r_{200}$. There are SPARC galaxies with best fit $M_\mathrm{b}>M_{200}$ and $r_\mathrm{e}>r_{200}$. Those are galaxies where the measured total rotation curves are always dominated by the baryonic component. None of those DM deficient galaxies will be used for SIDM cross section constraints.\par

\par

\section{Method}\label{sec:MCMC}
We fit the gravothermal fluid and isothermal solutions including the contribution of baryons to the gravitational potential (as derived in Section~\ref{sec:targetSelection}) to the SPARC rotation curves and BCG line-of-sight velocity dispersion measurements for the selected 9 baryon-poor systems. We explain how we model the BCG line-of-sight velocity dispersion and the observational effects in Appendix~\ref{apdx:vlos}. For all selected SPARC samples we solve the fluid formalism numerically with 150 log radial bins distributed uniformly in the range $-2\leq\log\hat{r}\leq3$\footnote{In this work $\log$ denotes a base-10 logarithm.}, while for galaxy clusters we grid the halo into 180 radial bins in range $-3\leq\log\hat{r}\leq3$. The time step is adaptively chosen such that the specific energy $\hat{u}=3\hat{v}_\mathrm{rms}^2/2$ changes by no more than 0.1\% among all the radial bins. We have tested that moving the inner boundary from $\log\hat{r}=-2$ to $\log\hat{r}=-3$ has negligible impact on the gravothermal solutions, but only reduces the time steps and slows down the halo evolution simulation process. We choose a wider radial binning range for BCGs so that the gravothermal solution radial extent covers all line-of-sight velocity dispersion data points. We solve the isothermal model over a sequence of halo evolution times such that the halo central velocity dispersion do not vary by more than 10\% over two adjacent time steps. All gravothermal and isothermal solutions are derived with a constant cross section model. 
We then apply the mapping method proposed by \defcitealias{2022arXiv220502957Y}{Yang2022}\citetalias{2022arXiv220502957Y} for estimating the gravothermal/isothermal solutions under arbitrary cross section models (Eq~\ref{eq:Rutherford_vis}-\ref{eq:Moller_vis}) and halo evolution time $t$.\par
\begin{figure}
    \centering
    \includegraphics[width=0.5\textwidth]{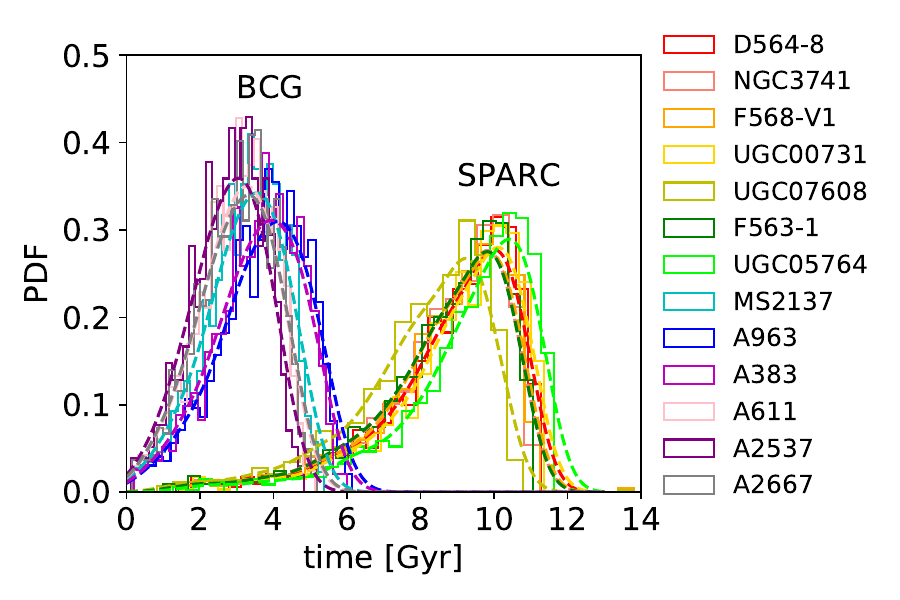}
    \caption{Halo evolution time PDFs for 7 baryon-poor SPARC galaxies and all BCGs. The histograms show PDFs of $t$ estimated by \textsc{Galacticus}, while the dashed curves show the best fit Gaussian KDEs, which are adopted as halo evolution time priors for the MCMC fits.}\label{fig:t_PDF}
\end{figure}

In principle the fits with either the gravothermal (isothermal) model for each SPARC galaxy should be processed over a 6 (5)-dimensional parameter space. For the gravothermal model the parameter space is $\{\rho_\mathrm{s},r_\mathrm{s},\sigma_0/m_\chi,\omega,\beta,t\}$, while for the isothermal model the parameter space is $\{\rho_\mathrm{s},r_\mathrm{s},\sigma_0/m_\chi,\omega,t\}$. When we fit SIDM models to the BCG cluster measurements we also consider variation of the stellar orbital anisotropy $\epsilon$, which brings an additional free parameter. At each point in the multi-dimensional parameter space we can compute the rotation curve or line-of-sight velocity dispersion radial profile from the mapped gravothermal/isothermal solution of each system. We thus perform a Monte Carlo Markov Chain (MCMC) analysis to derive constraints on cross section model parameters with \textsc{EMCEE} \citep{2013PASP..125..306F}. For all SPARC galaxies we set flat priors for parameters $\{\rho_\mathrm{s},r_\mathrm{s},\sigma_0/m_\chi,\omega,\beta\}$ as specified in Table~\ref{tb:MCMC_prior}. For galaxy clusters we use the same flat priors for $\sigma_0/m_\chi$, $\omega$, and $\beta$, but adopt epsilon-skew-normal (ESN) priors for $r_{200}$, $c_{200}$, and $M_{200}$ reported in \cite{2013ApJ...765...24N} to constrain $\rho_\mathrm{s}$ and $r_\mathrm{s}$. We set a flat prior $-0.2\leq\epsilon\leq0.2$ for the stellar velocity dispersion anisotropy, which is suggested by individual galaxy measurements \citep{2013ApJ...765...24N}. We notice that the gravothermal and isothermal solutions for all SPARC galaxies and clusters are derived under fixed $\{\rho_\mathrm{s},r_\mathrm{s}\}$, but we treat $\{\rho_\mathrm{s},r_\mathrm{s}\}$ as free parameters for SIDM cross section fits. This is not self-consistent because varying $\{\rho_\mathrm{s},r_\mathrm{s}\}$ effectively alters the baryonic gravitational potential in Eq~\ref{eq:2}. As discussed in Section~\ref{sec:model}, our method is justified because we have selected only baryon-poor systems. Altering or even completely ignoring the baryonic component in the selected systems has limited impact on the SIDM cross section fitting results. We therefore treat $\{\rho_\mathrm{s},r_\mathrm{s}\}$ as free parameters to avoid underestimating uncertainties in the parameters of the cross section model. The cross section model parameter $\sigma_0/m_\chi$ is degenerate with the halo evolution time. Given the halo mass $M_{200}$ at the observed redshift, we use the semi-analytic model \textsc{Galacticus} \citep{2012NewA...17..175B} to generate 1000 extended Press–Schechter based merger trees \citep{1974ApJ...187..425P,1991ApJ...379..440B,1991MNRAS.248..332B,1994MNRAS.271..676L,2008MNRAS.383..557P} with mass resolution of $10^{-3}M_{200}$. We then define the halo evolution time $t$ as the time for the main progenitor to grow from $M_{200}/2$ to $M_{200}$. We fit the halo evolution time distribution with a kernel-density estimate (KDE) using Gaussian kernels, and use the KDE probability distribution function (PDF) as the prior for the halo evolution time. The KDE PDFs of $t$ for the 7 selected SPARC galaxies and all BCGs are presented in Figure~\ref{fig:t_PDF}. The halo evolution time PDFs for SPARC galaxies and BCGs peak at 9--11 Gyr and 2--4 Gyr, respectively. This simply reflects the hierarchical structure formation of the Universe, where larger astronomical objects tend to form later.  We find increasing the EPS merger tree sample volume or mass resolution has negligible impact on the simulated PDF of $t$.\par

\begin{table}
\centering
\begin{tabular}{|c|c|}
\hline
$\log\sigma_0/m/[\mathrm{cm^2/g}]$&(-2.4,\ 4.0)\\ \hline
$\log\omega/[\mathrm{km/s}]$&(0.0,\ 4.0)\\ \hline
$\log\rho_\mathrm{s}/[M_\odot/\mathrm{kpc}^3]$&(3.0,\ 8.0)\\ \hline
$\log r_\mathrm{s}/[\mathrm{kpc}]$&(-1.0,\ 4.0)\\ \hline
$\beta$&(0.5,\ 1.5)\\ \hline
\end{tabular}
\caption{Summary of uniform prior bounds used in SIDM cross section model fittings with the SPARC galaxies. Parameter $\beta$ is only used for SIDM cross section fittings with the gravothermal solutions.}\label{tb:MCMC_prior}
\end{table}

\par
Although each system shows different scales and evolution times, the aim of this work is to find SIDM cross section models Eq~\ref{eq:Rutherford_vis}-\ref{eq:Moller_vis} that explain the DM density radial profiles for all systems. Taking the gravothermal fit as an example, a joint MCMC fit should contain in total $4\times7+5\times2+2=40$ free parameters. Here $4\times7$ refers to the unique $\theta=\{\rho_\mathrm{s},r_\mathrm{s},\beta,t\}$ parameters for each of the 7 selected SPARC galaxies. $5\times2$ stands for the unique $\theta=\{\rho_\mathrm{s},r_\mathrm{s},\beta,t,\epsilon\}$ parameters for the 2 selected galaxy clusters. $2$ corresponds to the cross section model parameter $\{\sigma_0/m_\chi,\omega\}$ that should be shared by all the SIDM halos. Since SPARC galaxies and galaxy clusters are independent astronomical objects, we can write the likelihood function as:
\begin{equation}
\mathcal{L}(X_{i=1\sim9}|\sigma_0/m_\chi,\omega,\theta_{i=1\sim9})=\prod_{i=1}^{9}\mathcal{L}(X_i|\sigma_0/m_\chi,\theta_i)\,.
\end{equation}
Here $X_i$ is the rotation curve or line-of-sight velocity dispersion radial profile measurement data for the $i^\mathrm{th}$ selected system. It is easy to show that the marginalized likelihood $\mathcal{L}(X_{i=1\sim9}|\sigma_0/m_\chi,\omega)$ given by a 40 parameter MCMC fit for all systems is identical to the product of marginalized likelihood $\mathcal{L}(X_i|\sigma_0/m_\chi,\omega)$ given by the MCMC fit for each system:

\begin{equation}
    \begin{split}
        &\mathcal{L}(X_{i=1\sim9}|\sigma_0/m_\chi,\omega)\\
        =&\int\displaylimits_{i=1\sim9} p(\theta_i) d\theta_i\mathcal{L}(X_{i=1\sim9}|\sigma_0/m_\chi,\omega,\theta_{i=1\sim9})\\
        =&\prod_{i=1}^{9}\int p(\theta_i)d\theta_i\mathcal{L}(X_i|\sigma_0/m_\chi,\omega,\theta_i)\\
        =&\prod_{i=1}^{9}\mathcal{L}(X_i|\sigma_0/m_\chi,\omega)\,.
    \end{split}
\end{equation}       
Here $p(\theta_i)$ is the posterior of parameter $x$ for the $i^{th}$ selected system. We assume Gaussian likelihood for each selected system, assuming the measured galaxy rotation curve or line-of-sight velocity dispersion data points among different radial bins are uncorrelated. This assumption is probably not
fully accurate for the SPARC galaxies where the points may be correlated due
to the beam-smearing effects. An improved analysis could attempt to estimate the
correlation matrix of the rotation curve data points and further compute a more accurate likelihood.

\section{Results}\label{sec:result}
\begin{figure*}
    \centering
    \includegraphics[width=0.7\textwidth]{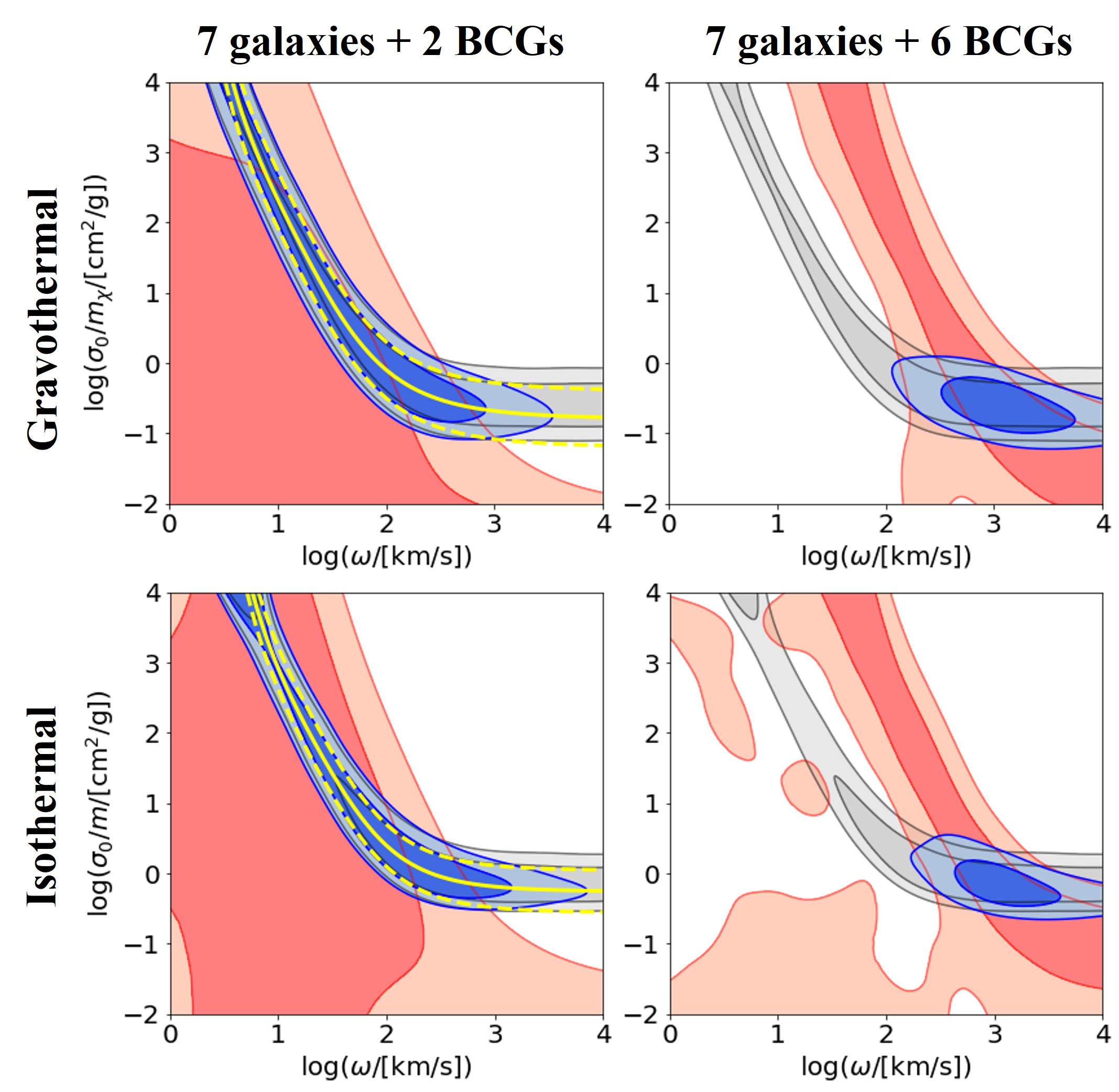}
    \caption{SIDM cross section fitting results for model Eq~\ref{eq:Rutherford_vis} at 68\% and 95\% confidence levels. The first (second) row show constraints derived from the gravothermal (isothermal) model. In each panel, the grey posterior is contributed by 7 SPARC galaxies. The red posterior is contributed by BCGs. The blue contours show galaxy-BCG joint constraints. In the first column, the BCG constraints are only provided by the baryon-poor A2667 and A2390, while for the second column the BCG constraints are provided by 6 relaxed galaxy clusters. Yellow solid and dashed curves in the first column show the best fit double power law $\log\sigma_0/m-\log\omega$ relation and its $1\sigma$ uncertainty, as summarized in Eq~\ref{eq:Rutherford_gravothermal_bestfit}-\ref{eq:Rutherford_isothermal_bestfit}.}\label{fig:SIDMfit_rutherford}
\end{figure*}

\begin{figure*}
    \centering
    \includegraphics[width=0.7\textwidth]{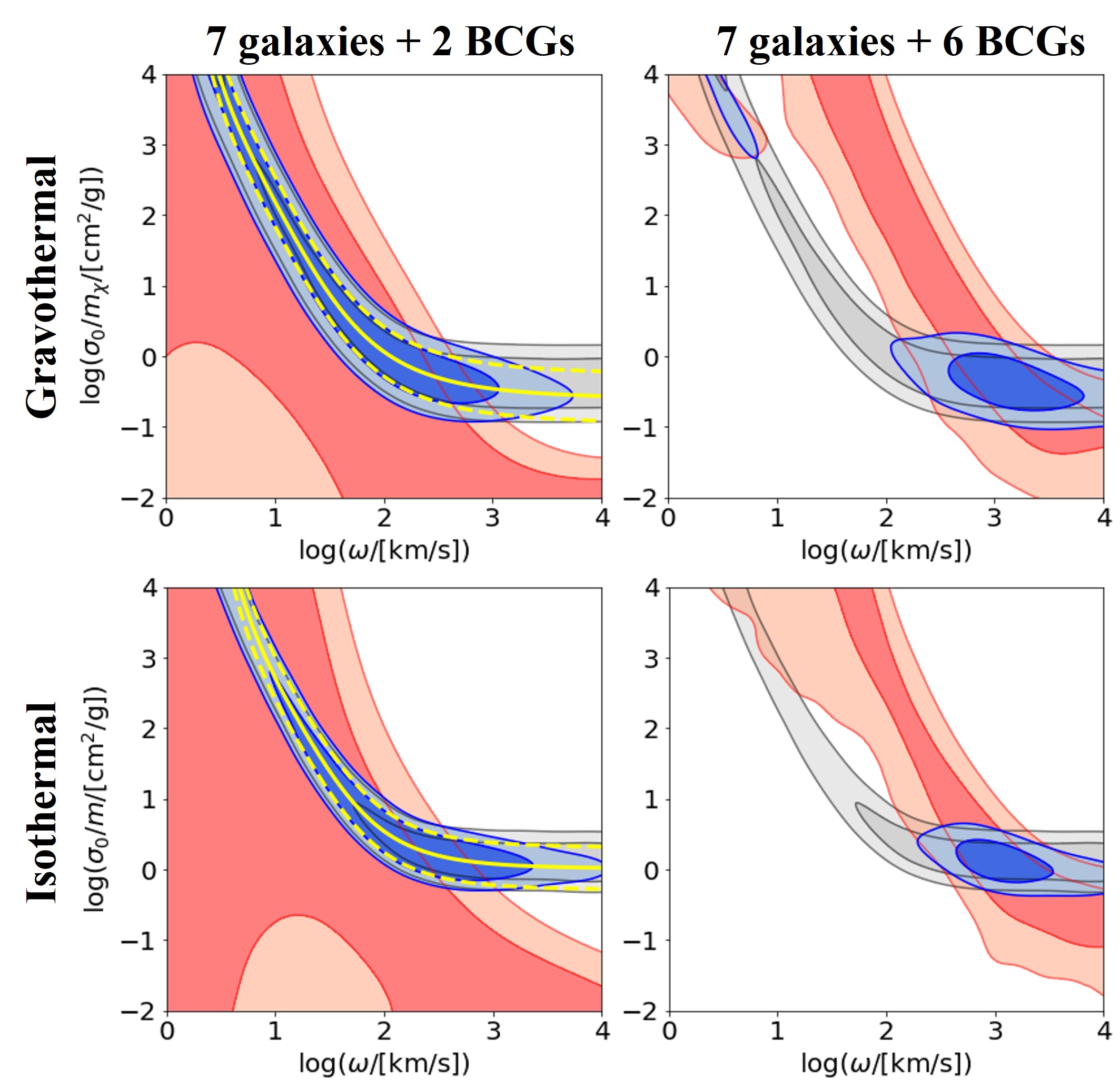}
    \caption{This figure is similar to Figure~\ref{fig:SIDMfit_rutherford}, but show SIDM cross section fitting results for model Eq~\ref{eq:Moller_vis} at 68\% and 95\% confidence levels.}\label{fig:SIDMfit_Moller}
\end{figure*}

Results for $\sigma_0/m_\chi-\omega$ fitting marginalized over all the halo specific parameters constrained jointly by the 9 selected systems are presented in the first column of Figure~\ref{fig:SIDMfit_rutherford} for the Rutherford scattering case and Figure~\ref{fig:SIDMfit_Moller} for the M{\o}ller scattering case. In Figure~\ref{fig:SIDMfit_rutherford}-\ref{fig:SIDMfit_Moller}, the top row shows constraints derived from the gravothermal solutions, while the bottom row show constraints from the isothermal model. In each panel, the SIDM cross section model parameter space favored by the 7 SPARC galaxies is presented in the ``L"-shaped grey band, turning at characteristic velocity $\omega\sim100$ km/s. This ``L"-shaped parameter degeneracy is caused by the asymptotic features of Eq~\ref{eq:Rutherford_vis}-\ref{eq:Moller_vis}. Taking the cross section model for Rutherford scattering Eq~\ref{eq:Rutherford_vis} as an example, at low relative velocity $v_{12}\ll\omega$, $\sigma_V\rightarrow\sigma_0$, while at high relative velocity $v_{12}\gg\omega$, $\sigma_V\rightarrow12\sigma_0\omega^4\ln(v_{12}/\omega)/v_{12}^4$. Since the central velocity dispersions of the selected SPARC galaxies are of order $\sim 100$ km/s, $\sigma_0/m_\chi$ is degenerate with $\sim\omega^4$ when $\omega\ll100$ km/s. On the other hand, the posterior of $\sigma_0/m_\chi$ becomes insensitive to $\omega$ variations for $\omega\gg 100$ km/s. Similarly, the parameter constrained by the selected 2 BCGs also show this ``L"-shaped degeneracy, although turning at a larger characteristic velocity $\omega\sim1000$ km/s, as shown in the red 2D posterior. However, since DM density radial profiles of the 2 selected BCGs are consistent with NFW, corresponding to a negligible SIDM cross section, the BCGs contribute only an upper limit on the SIDM cross section. The above discussion also applies to the M{\o}ller scattering cross section model Eq~\ref{eq:Moller_vis}, although at $v_{12}\ll\omega$, $\sigma_V\rightarrow0.5\sigma_0$. This factor of 2 difference is caused by the fact that the two particles involved in the M{\o}ller scattering process are indistinguishable, and a scattering angle of $\theta$ or $\pi-\theta$ are identical events. This slightly different asymptotic behavior causes M{\o}ller scattering constraints on $\sigma_0/m_\chi$ to be greater than those derived under the Rutherford scatter by a factor of 2 at large $\omega$. At $\omega\gtrsim1000$ km/s, the constrained cross section models are essentially velocity independent, i.e. $\sigma=\sigma_0$, for all the selected galaxies and BCGs. The constant cross section constraints are broadly in agreement with those reported by other works in the literature : galaxies with lower characteristic velocities show SIDM cross sections of $\sim 1$ cm$^2$/g, while clusters generally show smaller cross sections $\sim0.1$ cm$^2$/g \citep{2016PhRvL.116d1302K,2018ApJ...853..109E,2021JCAP...01..024S,2021MNRAS.503..920C,2022MNRAS.510...54A}. The SIDM model parameter space constrained jointly by the 7 SPARC galaxies and 2 BCGs is presented in the blue posterior. Although current data cannot strongly constrain the SIDM cross section model, we are nevertheless able to extract a best fit $\{\sigma_0/m_\chi,\omega\}$ relation for Rutherford scattering model:
\begin{equation}\label{eq:Rutherford_gravothermal_bestfit}
\begin{split}
&\log\left(\dfrac{\sigma_0/m_\chi}{[\mathrm{cm^2/g}]}\right)\\
=&\dfrac{2.42}{\left(\dfrac{\log(\omega/[\mathrm{km/s}])}{1.70}\right)^{0.648}+\left(\dfrac{\log(\omega/[\mathrm{km/s}])}{1.70}\right)^{5.17}}-0.799\,,
\end{split}
\end{equation}
with $1\sigma$ scatter of $\sim0.4$ dex and $\log(\omega/[\mathrm{km/s}])\leq2.9$ $1\sigma$ upper bound through the gravothermal fluid method.\par 
The best fit $\{\sigma_0/m_\chi,\omega\}$ relation for Rutherford scatterings constrained by the isothermal model is:
\begin{equation}\label{eq:Rutherford_isothermal_bestfit}
\begin{split}
&\log\left(\dfrac{\sigma_0/m_\chi}{[\mathrm{cm^2/g}]}\right)\\
=&\dfrac{1.80}{\left(\dfrac{\log(\omega/[\mathrm{km/s}])}{1.85}\right)^{0.981}+\left(\dfrac{\log(\omega/[\mathrm{km/s}])}{1.85}\right)^{6.48}}-0.260\,,
\end{split}
\end{equation}
with $\sim0.3$ dex of $1\sigma$ scatter and $\log(\omega/[\mathrm{km/s}])\leq3.2$ at 68\% confidence level. We show the best fit $\{\sigma_0/m_\chi,\omega\}$ double power-law relation and its scatter at 68\% confidence levels as the yellow solid and dashed curves in each panel.\par 
The best fit $\{\sigma_0/m_\chi,\omega\}$ relation for M{\o}ller scattering constrained by the gravothermal fluid formalism is:
\begin{equation}
\begin{split}
&\log\left(\dfrac{\sigma_0/m_\chi}{[\mathrm{cm^2/g}]}\right)\\
=&\dfrac{2.57}{\left(\dfrac{\log(\omega/[\mathrm{km/s}])}{1.60}\right)^{0.498}+\left(\dfrac{\log(\omega/[\mathrm{km/s}])}{1.60}\right)^{4.43}}-0.606\,,
\end{split}
\end{equation}
with $1\sigma$ scatter of $\sim0.35$ dex and $\log(\omega/[\mathrm{km/s}])\leq3.1$ $1\sigma$ upper bound. M{\o}ller scattering cross section model constrained by the isothermal method is:
\begin{equation}
\begin{split}
&\log\left(\dfrac{\sigma_0/m_\chi}{[\mathrm{cm^2/g}]}\right)\\
=&\dfrac{1.76}{\left(\dfrac{\log(\omega/[\mathrm{km/s}])}{1.74}\right)^{0.845}+\left(\dfrac{\log(\omega/[\mathrm{km/s}])}{1.74}\right)^{5.85}}+0.00972\,,
\end{split}
\end{equation}
with $1\sigma$ scatter of $\sim0.3$ dex and $\log(\omega/[\mathrm{km/s}])\leq3.4$ $1\sigma$ upper bound. This degenerate model constraint will provide quantitative guidance for upcoming SIDM simulations and lensing survey forecasts \cite[e.g.][]{2020ApJ...896..112N,2021MNRAS.507.2432G,2022MNRAS.513.4845Z}. The SIDM cross section constraints derived from the isothermal approach tend to be tighter than those derived from the gravothermal model. This is because the gravothermal fits contain an additional free parameter $0.5\leq\beta\leq1.5$, which is degenerate with the cross section. Overall, the SIDM cross section constraints provided by the gravothermal and isothermal models are consistent within $1\sigma$ confidence level, but $\sigma_0/m_\chi$ preferred by the gravothermal solutions are smaller than those constrained by the isothermal solutions by a factor of $\sim3$.\par

We notice that the target selection criterion introduced in Section~\ref{sec:targetSelection} might be too conservative. Our goal is to select systems in which the baryon gravitational potential does not alter the SIDM halo gravothermal evolution significantly. However, the specific definition of what constitutes a `baryon-poor' object in this context requires further study and calibration using cosmological simulations\footnote{For example, it may be possible to use simulations to determine a criterion on the mass of baryons, or their gravitational potential at some characteristic radius (e.g. $r_\mathrm{s}$) which ensures that the gravothermal solution remains within some small window around the baryon-free solution.}. Moreover, the main motivation for us to select baryon-poor system is to limit the impacts of the varying baryonic potential during the MCMC fits, which is less of a problem for the BCGs due to the large radial extend of the measurements. If we loosen the BCG selection criterion such that we include all galaxy clusters measured in \cite{2013ApJ...765...24N} besides A2537 for SIDM cross section model constraint, we find the joint posterior presented in Figure~\ref{fig:SIDMfit_rutherford}-\ref{fig:SIDMfit_Moller} second column. We exclude A2537 here because this BCG is likely disturbed and shows a multi-component structure. We find MS2137, A963, and A383 show dark matter density profiles with cored regions at the halo center, therefore ruling out CDM and contributing a lower SIDM cross section limit. This larger BCG measurement data set helps to pin down the SIDM cross section model with best fit parameters summarized in Table~\ref{tb:bestfit_param_summary}.\par 
 
\par

\par

\begin{table}
\centering
\begin{tabular}{|l|c|c|}
\hline
&gravothermal soln&isothermal soln\\ \hline
Rutherford&$-0.6\pm0.4\,,3.1\pm0.6$ &$-0.2^{+0.4}_{-0.3}\,,3.0^{+0.6}_{-0.4}$\\ \hline
M{\o}ller&$-0.4\pm0.4\,,3.1_{-0.5}^{+0.7}$ &$0.1\pm0.3\,,3.0^{+0.5}_{-0.4}$\\ \hline
\end{tabular}
\caption{Best fit SIDM cross section model parameters constrained by 7 galaxies and 6 BCGs with different halo evolution models. The first (second) value of each cell specifies the 1$\sigma$ constraints for $\sigma_0/m_\chi/[\mathrm{cm^2/g}]$ ($\omega/[\mathrm{km/s}]$). }\label{tb:bestfit_param_summary}
\end{table}

\begin{figure*}
    \centering
    \includegraphics[width=1\textwidth]{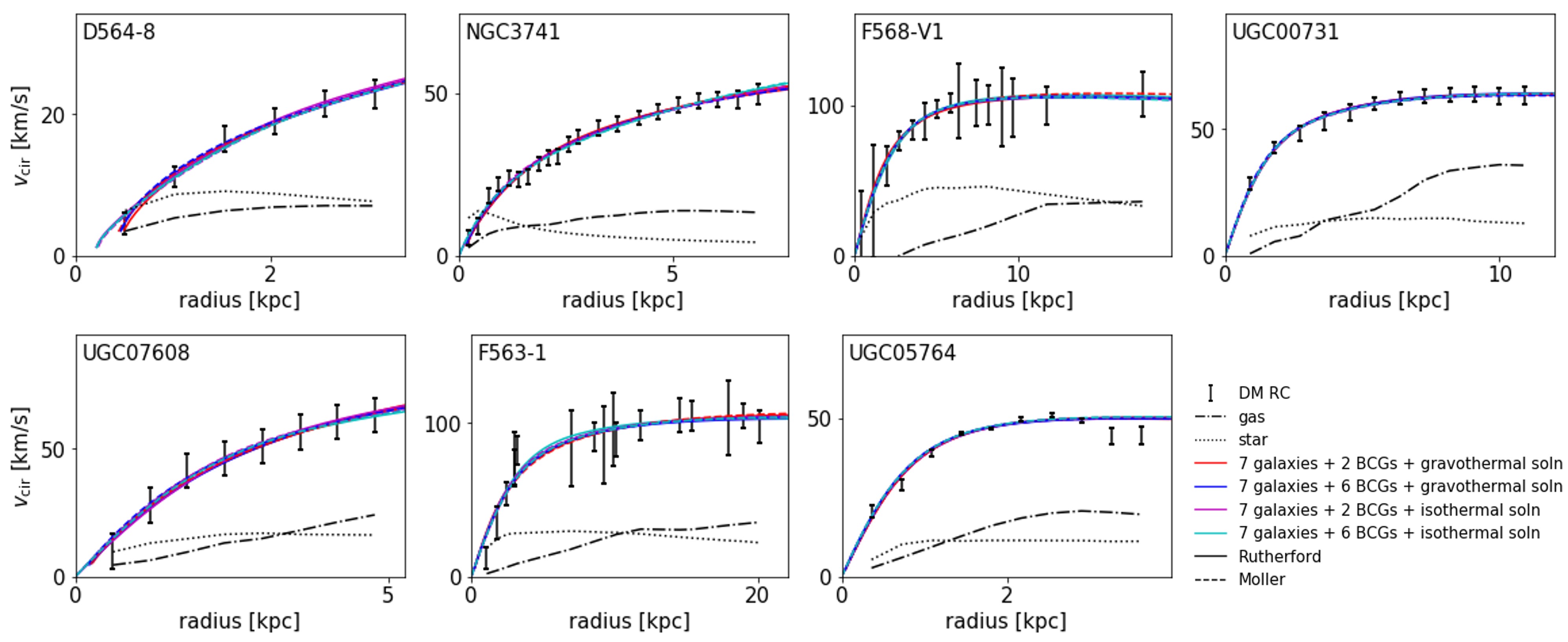}
    \caption{Comparison between SPARC rotation curves \citep{2016AJ....152..157L} and the example best-fit gravothermal solutions among the 7 selected baryon-poor galaxies. The example best-fit SIDM cross section models are specified in Table~\ref{tb:bestfit_param}. The best-fit SIDM cross section models derived from different data sets and different SIDM halo evolution models describe all the SPARC rotation curves well.}\label{fig:SPARC_bestfit}
\end{figure*}

\begin{figure*}
    \centering
    \includegraphics[width=1\textwidth]{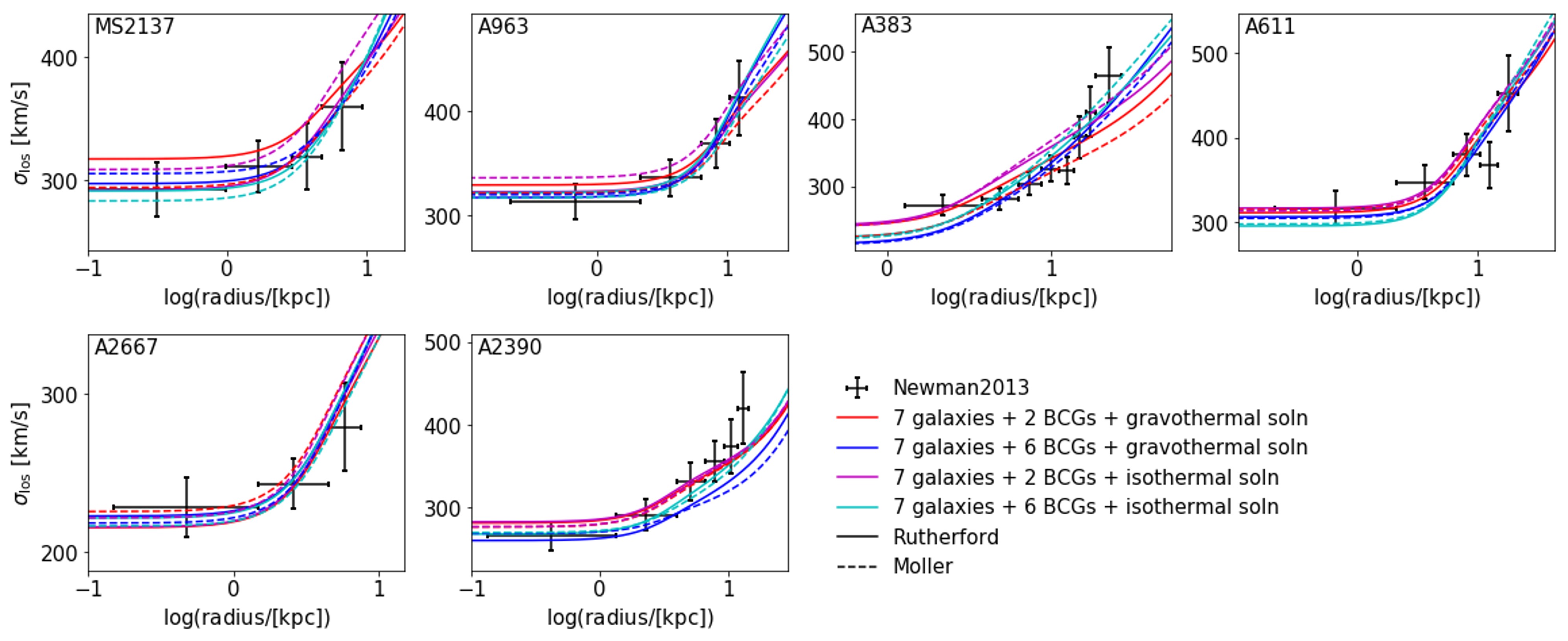}
    \caption{Comparison between BCG line-of-sight velocity dispersion measurements \citep{2013ApJ...765...24N} and the best-fit gravothermal solutions among 6 relaxed BCGs. The best-fit curve notations are identical with Figure~\ref{fig:SPARC_bestfit}. SIDM cross section model constrained by 7 SPARC galaxies and 2 BCGs (presented in red and magenta curves) describe measurements for A2667 and A2390 better than the blue and cyan models, which are constrained jointly by 7 galaxies and 6 BCGs. However, the blue and cyan models reproduce measurements for other BCGs, especially A383, better than models presented by the red and magenta curves. Since baryonic components in the other 4 BCGs can significantly alter the halo gravothermal evolution, in this work we do not claim the more stringent dashed best fit results as reliable.}\label{fig:BCG_bestfit}
\end{figure*}

In Figure~\ref{fig:SPARC_bestfit} and Figure~\ref{fig:BCG_bestfit} we show example comparisons between the best fit model predictions and observations, although there is no single best fit SIDM cross section model preferred by the selected 9 baryon-poor systems. Specifically, we select $\{\log\sigma_0/m,\log\omega\}$ in the parameter space that corresponds to the maximum likelihood in the galaxy-BCG joint fits for each of the four cases presented in Figure~\ref{fig:SIDMfit_rutherford}-\ref{fig:SIDMfit_Moller}. We summarize the selected best fit $\{\log\sigma_0/m,\log\omega\}$ combinations in Table~\ref{tb:bestfit_param}, but we emphasize that those are not the only best fit SIDM cross section models preferred by the selected 7 galaxies and 2 BCGs. In Figure~\ref{fig:SPARC_bestfit} (Figure~\ref{fig:BCG_bestfit}) we compare the measured SPARC rotation curves (BCGs line-of-sight velocity dispersion radial profiles) with the example best fit results. The halo evolution time $t$, $\beta$ parameter for the gravothermal model, $\rho_\mathrm{s}$ and $r_\mathrm{s}$ of each system are set as the best fit values marginalized over the fixed $\{\log\sigma_0/m,\log\omega\}$. We confirm in Figure~\ref{fig:SPARC_bestfit} that those example best fit SIDM cross section models can reproduce all of the SPARC rotation curves well. Figure~\ref{fig:BCG_bestfit} shows that the example best fit cross section models presented in red and magenta curves better reproduce the measured line-of-sight velocity dispersion for BCG A2667 and A2390 than the dashed models. However, the blue and cyan models better describe the projected velocity dispersion measurements for the other four BCGs, especially for BCG A383. As we have discussed in Section~\ref{sec:targetSelection}, we prefer to remain conservative about the baryonic effects in MS2137, A963, A383, and A611, that could significantly alter the halo gravothermal evolution. Therefore, in this work we do not claim the more stringent fit presented by the dashed curves as reliable.\par

\begin{table}
\centering
\begin{tabular}{|l|c|c|}
\hline
&$\log\sigma_0/m_\chi$&$\log\omega$\\ \hline
7 galaxies, 2 BCGs, gravothermal soln&-0.13/-0.11&2.0/2.2\\ \hline
7 galaxies, 6 BCGs, gravothermal soln&-0.62/-0.36&3.1/3.1\\ \hline
7 galaxies, 2 BCGs, isothermal soln&-0.024/0.20&2.5/2.5\\ \hline
7 galaxies, 6 BCGs, isothermal soln&-0.17/0.11&3.0/3.0\\ \hline
\end{tabular}
\caption{Example best fit SIDM cross section model parameters derived from different observational data sets and SIDM halo evolution models. The first/second value in each cell corresponds to example best fit result under Rutherford/M{\o}ller scattering. $\sigma_0/m_\chi$ is in units of cm$^2$/g, while $\omega$ is in units of km/s. Since the selected 2 baryon-poor BCGs cannot break the $\sigma_0/m_\chi-\omega$ degeneracy, there is no single best fit SIDM cross section model constrained by the 7 SPARC galaxies and 2 BCGs.}\label{tb:bestfit_param}
\end{table}

\par

\par

\section{Conclusion and discussion}\label{sec:discuss}
In this work we constrain physically motivated SIDM cross section models (Eq~\ref{eq:Rutherford_vis}-\ref{eq:Moller_vis}) through fitting gravothermal fluid and isothermal solutions to DM halo profile measurements. To break the degeneracy of the two free parameters $\{\sigma_0/m_\chi,\omega\}$ introduced by Eq~\ref{eq:Rutherford_vis}-\ref{eq:Moller_vis}, we constrain the model with DM profile measurements from two classes of astrophysical systems that show very different central velocity dispersion: galaxies and BCGs. We select 7 SPARC galaxies and 2 BCGs that are isolated and baryon-poor, so that tidal effects and the gravitational potential of the baryonic component make negligible impact on the SIDM halo gravothermal evolution.\par 
For each system we perform MCMC parameter fits for the gravothermal and isothermal solutions to account for SIDM cross section model uncertainties contributed by possible $\rho_\mathrm{s},r_\mathrm{s},\beta, \epsilon$ and halo evolution time variations. The efficiency of the gravothermal fluid formalism is optimized through a mapping method proposed by \defcitealias{2022arXiv220502957Y}{Yang2022}\citetalias{2022arXiv220502957Y}, so that the gravothermal solutions can be rapidly estimated in the continuous multi-dimensional parameter space. The isothermal method is usually believed to be only valid for describing SIDM halo evolution during its core formation stage under a constant cross section. In this work we prove that the high density isothermal solutions describe the SIDM halo distribution during its core collapse phase. We also prove that the mapping method introduced in \defcitealias{2022arXiv220502957Y}{Yang2022}\citetalias{2022arXiv220502957Y} is applicable to the isothermal model, assuming the isothermal and gravothermal solutions share similar evolution universality. We are therefore able to extend the isothermal model to the full SIDM halo evolution process under arbitrary velocity dependent cross section models.\par 
We find the two selected BCGs show DM density profiles consistent with NFW, resulting in failure to fully break the $\sigma_0/m_\chi-\omega$ parameter degeneracy, instead resulting only in an SIDM cross section upper limit. Combining the BGCs with the low-mass galaxy rotation curves, We report the degenerate best fit relations, $\omega$ upper bounds, and $1\sigma$ scatters for the gravothermal and isothermal fits. The isothermal constraints are slightly tighter than the gravothermal constraints since the gravothermal fits contain an additional free parameter $0.5\leq\beta\leq1.5$. SIDM cross sections preferred by the gravothermal and isothermal solutions are in agreement with each other at the $1\sigma$ confidence level, but the constraints provided by the isothermal model are tighter than the gravothermal constraints by a factor of $\sim3$. These degenerate best fit results will be useful for upcoming SIDM simulations and survey forecasts. More stringent cross section constraints may be achieved with current DM distribution measurements if detailed fluid formalism versus idealized SIDM simulation calibration is performed, accounting for the presence of large baryonic components that can significantly alter the SIDM halo evolution.\par
In this work we do not quantify the influences of the mass-to-light ratio parameter $\gamma_*$ on the SIDM cross section constraints. A larger $\gamma_*$ corresponds to a deeper baryonic gravitational potential well and, therefore, larger baryonic effect uncertainties. It will also make our MCMC fitting strategy less self-consistent, as discussed in Section~\ref{sec:targetSelection}. If we were to introduce $\gamma_*$ as an additional free parameter in the MCMC fitting process for the gravothermal fluid formalism in a self-consistent way, it would require re-solving the gravothermal PDEs every time a new $\gamma_*$ is proposed by the MCMC algorithm. This breaks the advantage of the gravothermal solution mapping method used in this work and will be too computationally expensive. It is also challenging to solve the extended isothermal model introduced in Section~\ref{subsec:iso} with varying $\gamma_*$ in every MCMC step since we would need to find the low-density solution, high-density solution, and $t_\mathrm{merge}$ simultaneously. Solving for these at every MCMC step would significantly slow down the fitting process. A more practical way to achieve this would be to create an isothermal solution lookup table for a range of baryonic profiles, and we will explore this option in future work. Qualitatively, for a fixed total  radial mass profile for a given system, a larger $\gamma_*$ generally corresponds to a more cored halo and a larger SIDM cross section. We have tested that adopting baryonic component density profiles to be $1\sigma$ larger or smaller than the bestfit values for all BCGs can influence the upper bound of the SIDM cross section constraints by a factor of 3 to 4. \par

\section{Acknowledgements}
We thank Andrew Newman for instructions about modeling the observational effects in the BCG line-of-sight velocity dispersion profiles. We thank Haibo Yu and Daneng Yang for useful discussions.  This work was supported in part by the NASA Astrophysics Theory Program, under grant 80NSSC18K1014. MSF is funded by the \emph{Deutsche Forschungsgemeinschaft (DFG, German Research Foundation)} under Germany’s Excellence Strategy -- EXC-2094 ``Origins'' -- 390783311.

\section*{Data availability}
The data used to support the findings of this study are available from the corresponding author upon request.

\appendix
\section{Hernquist fits for baryonic components among the 9 selected systems}\label{apdx:Hernquist}
\begin{figure*}
    \centering
    \includegraphics[width=1\textwidth]{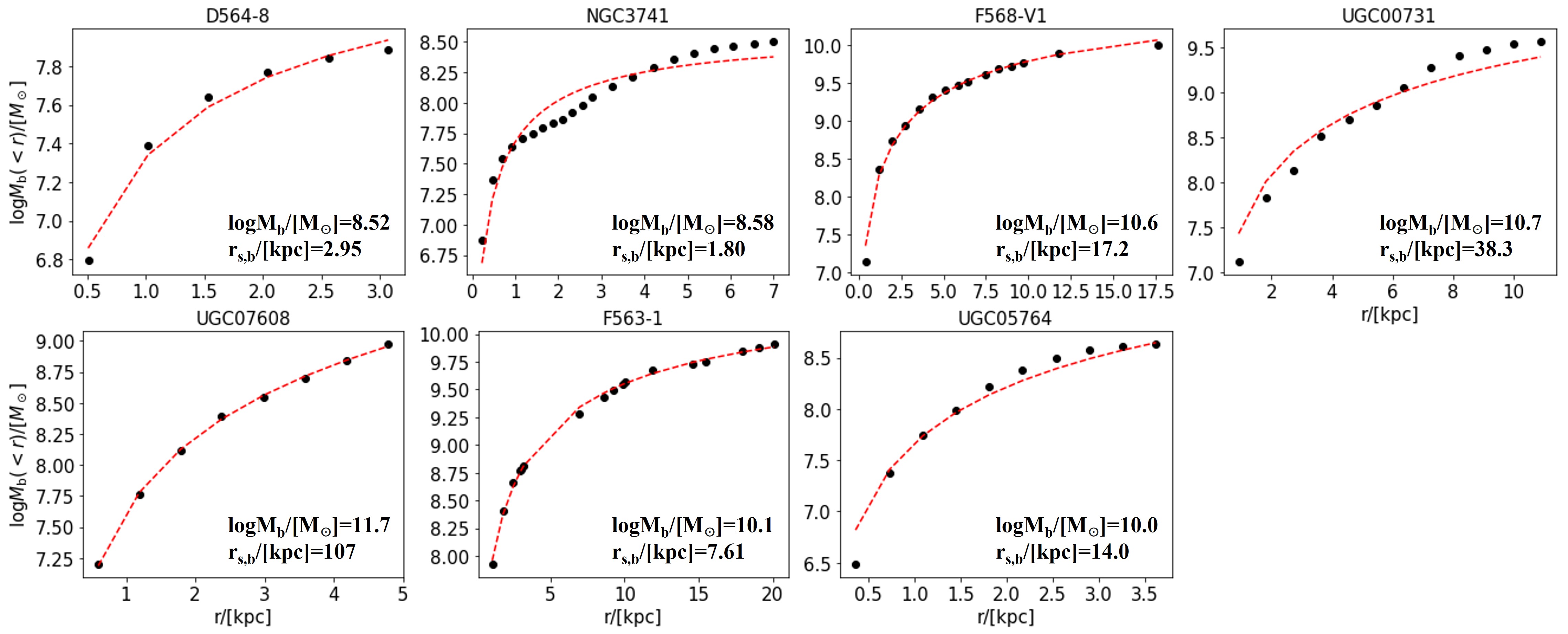}
    \caption{Comparison between the measured baryonic enclosed mass and the best fit Hernquist profile for the 7 baryon-poor SPARC galaxies. The measurements are presented by black scatters, while the best fit results are shown by the red dashed curves.}\label{fig:Hernquist_SPARC}
\end{figure*}

\begin{figure*}
    \centering
    \includegraphics[width=1\textwidth]{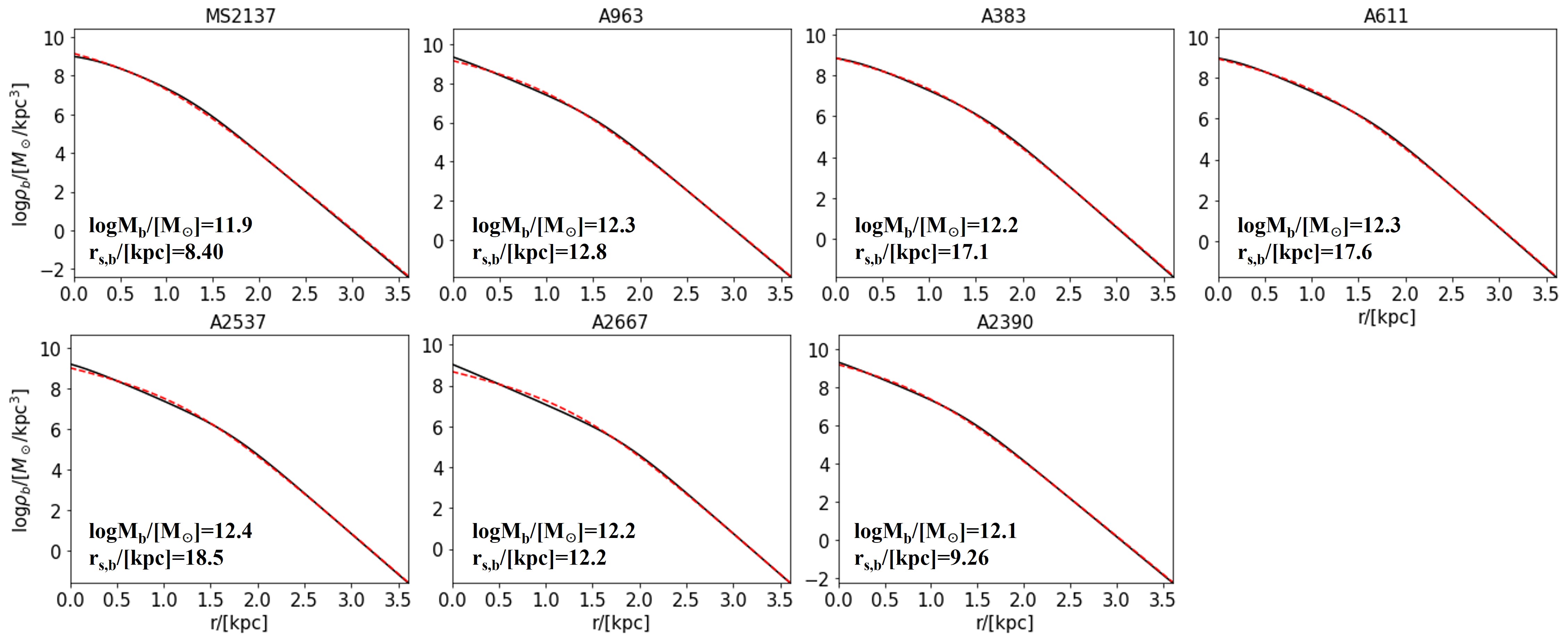}
    \caption{Comparison between the best fit dPIE profile and Hernquist profile for 7 BCGs. The dPIE profiles are shown in black solid curves, while the best fit Hernquist profiles are shown by red dashed curves.}\label{fig:Hernquist_BCG}
\end{figure*}

In this appendix we compare the best fit Hernquist profiles with SPARC baryonic rotation curves $v_\mathrm{b,cir}(r)$ and the best fit BCG luminosity tracer dPIE profiles for the selected 7 galaxies and 7 BCGs. Figure~\ref{fig:Hernquist_SPARC} compares the measured baryonic enclosed mass $M_\mathrm{b}(<r)=v_\mathrm{b,cir}^2r/G$ (black dots) with the best fit Hernquist profile (red dashed). Figure~\ref{fig:Hernquist_BCG} presents comparison between the dPIE profiles for the BCG stellar mass distribution fitted by \cite{2013ApJ...765...24N} (black solid) with the best fit Hernquist profile (red dashed). The best fit Hernquist profile parameters for each system are specified in the corresponding panel.\par

\par

\par

\section{Line-of-sight velocity dispersion}\label{apdx:vlos}
In this appendix we explain the method used to model the observational effects in the line-of-sight velocity dispersion measurements. We also refer readers to \cite{2004ApJ...604...88S} for a similar discussion.\par
At each point in the multi-dimensional parameter space for the gravothermal or isothermal model fitting, we can derive the stellar 1D velocity dispersion radial profile $v_\mathrm{rms,b}(r)$ through the mapping method \defcitealias{2022arXiv220502957Y}{Yang2022}\citetalias{2022arXiv220502957Y} and the following Jeans equation:
\begin{equation}
    \dfrac{\partial(\rho_bv_\mathrm{rms,b}^2)}{\partial r}=-\dfrac{G(M+M_b)\rho_b}{r^2}-2\epsilon\dfrac{\rho_bv_\mathrm{rms,b}^2}{r}\,.
\end{equation}
Here $\rho_b$ is the stellar dPIE density profile fit by \cite{2013ApJ...765...24N}, and $\epsilon$ is the stellar orbital anisotropy. The projected velocity dispersion without observational effects $\sigma_\mathrm{p}(r)$ is \citep{2018NatAs...2..907V}:
\begin{equation}
\Sigma_b(r)\sigma_\mathrm{p}^2(r)=2\int_r^\infty\dfrac{r'dr'}{\sqrt{r'^2-r^2}}(1-\epsilon\dfrac{r^2}{r'^2})\rho_b(r')v_\mathrm{rms,b}^2(r')\,.
\end{equation}
Here $\Sigma_b$ is the surface density profile of the stellar tracers, derived from $\rho_b$ via an Abel transform:
\begin{equation}
    \Sigma_b(r)=2\int_r^\infty\dfrac{\rho_b(r')r'dr'}{\sqrt{r'^2-r^2}}\,.
\end{equation}\par
Before comparing $\sigma_\mathrm{p}(r)$ given by the gravothermal or isothermal model with measurements, we need to account for two observational effects. The first effect is astronomical seeing caused by turbulence in the Earth's atmosphere, which blurs the BCG images. The second factor is that BCG spectra are measured through a slit with finite width, so the radial binning is not strictly defined in a spherical coordinate. To model the above two observational effects, we assign $I(r)\sigma^2_\mathrm{p}(r)$ over a $1000\times1000$ 2D grid extending over the interval $-100\leq x/[\mathrm{kpc}]\leq100$ and $100\leq y/[\mathrm{kpc}]\leq100$. Here $r=\sqrt{x^2+y^2}$. $I(R)$ is the BCG  dPIE surface brightness profile. We convolve the $I(r)\sigma^2_\mathrm{p}(r)$ 2D image with a Gaussian seeing point-spread function. The Gaussian kernel full-width-half maximum for each BCG is provided by \cite{2013ApJ...765...24N}. We then mask out the $I(r)\sigma^2_\mathrm{p}(r)$ signal at $y\geq d_\mathrm{A}\theta/2$ and measure the binned $\langle I\sigma_\mathrm{p}^2\rangle(r)$ along the $x$ axis. Here $d_\mathrm{A}$ is the BCG angular diameter distance, and $\theta$ is the slit width. We repeat the above calculation to model $\langle\sigma_\mathrm{p}^2\rangle(r)$. Finally we estimate the line-of-sight velocity dispersion radial profile with observational effects as $\sigma_\mathrm{los}^2=\sqrt{\langle I\sigma_\mathrm{p}^2\rangle/\langle\sigma_\mathrm{p}^2\rangle}$.\par




\vspace{-5mm}
\bibliographystyle{mnras}
\bibliography{SIDMsigma}




\label{lastpage}
\end{document}